\documentclass[10pt]{sig-alternate-05-2015}
\usepackage[
  pass,%
]{geometry}
\usepackage{nohyperref}
\usepackage{flushend}
\usepackage{enumitem}
\usepackage{cite}
\usepackage{graphicx}
\usepackage{grffile}
\usepackage{amssymb}
\usepackage{amsmath}
\usepackage{multicol,multirow}
\usepackage{rotating,url}
\usepackage{enumitem}
\usepackage{times}
\usepackage{color}
\usepackage{subfigure}
\usepackage{xspace}
\usepackage{ifpdf}
\usepackage{mdwlist}
\usepackage{colortbl}
\usepackage{amssymb}
\usepackage{booktabs}
\usepackage{multirow}

\newcommand{\ie}{i.e., \@}
\newcommand{\eg}{e.g., \@}

\newcommand{\parax}[1]{\vspace{0.1em} \noindent \textbf{#1:}}
\newcommand{\para}[1]{\vspace{0.1em} \noindent \textbf{#1}}
\long\def\comment#1{}

\begin{document}

\CopyrightYear{2016}
\setcopyright{acmlicensed}
\conferenceinfo{IMC 2016,}{November 14 - 16, 2016, Santa Monica, CA, USA}
\isbn{978-1-4503-4526-2/16/11}\acmPrice{\$15.00}
\doi{http://dx.doi.org/10.1145/2987443.2987473}

\clubpenalty=10000 
\widowpenalty = 10000

\title{Beyond Counting:\\ New Perspectives on the Active IPv4 Address Space} %

\numberofauthors{3} %
\author{
\alignauthor\mbox{Philipp Richter}\\
     \affaddr{TU Berlin / Akamai}\\
    \affaddr{{prichter@inet.tu-berlin.de}}
\alignauthor \hspace{-0.4in}\mbox{Georgios Smaragdakis}\\
     \affaddr \hspace{-0.4in}{MIT}\\
    \affaddr \hspace{-0.4in}{{gsmaragd@csail.mit.edu}}
\alignauthor \hspace{-1.1in}\mbox{David Plonka}\\
     \affaddr \hspace{-1.1in}{Akamai}\\
     \affaddr \hspace{-1.1in}{{plonka@akamai.com}}
\alignauthor \hspace{-2in}\mbox{Arthur Berger}\\
     \affaddr \hspace{-2in}{Akamai / MIT }\\
     \affaddr \hspace{-2in}{{arthur@akamai.com}}
\vspace{1em}
}

\maketitle

\abstract{
In this study, we report on techniques and analyses that enable us to capture 
Internet-wide activity at individual IP address-level granularity by relying on 
server logs of a large commercial content delivery network (CDN) that serves 
close to 3 trillion  HTTP requests on a daily basis. Across the whole of 2015, 
these logs recorded client activity involving 1.2 billion unique IPv4 
addresses, the highest ever measured, in agreement with recent estimates. 
Monthly client IPv4 address counts showed constant growth for years prior, but 
since 2014, the IPv4 count has stagnated while IPv6 counts have grown. Thus, it 
seems we have entered an era marked by increased complexity, one in which the 
sole enumeration of active IPv4 addresses is of little use to characterize 
recent growth of the Internet as a whole.

With this observation in mind, we consider new points of view in the study of 
global IPv4 address activity. Our analysis shows significant churn in active 
IPv4 addresses: the set of active IPv4 addresses varies by as much as 25\% over 
the course of a year. Second, by looking across the active addresses in a 
prefix, we are able to identify and attribute activity patterns to network 
restructurings, user behaviors, and, in particular, various address assignment 
practices. Third, by combining spatio-temporal measures of address utilization 
with measures of traffic volume, and sampling-based estimates of relative host 
counts, we present novel perspectives on worldwide IPv4 address activity, 
including empirical observation of under-utilization in some areas, and 
complete utilization, or exhaustion, in others.

\section{Introduction}\label{sec:intro}

The Internet continuously evolves as a result of the interaction among different stakeholders with diverse {\em business
strategies}, {\em operational practices}, and access to {\em network 
resources}~\cite{Tussle:conf}. This evolution occurs in the context of the 
architecture of the Internet, which is based on a number of principles that 
guarantee basic connectivity and interoperability, including global addressing, 
realized with the Internet Protocol (IP). 
This has motivated a number of researchers to study address space utilization characteristics to assess the current state and expansion of the Internet~\cite{Capturing-Ghosts:IMC2014,Understanding-Block-Level-Address:SIGCOMM2010,Tinocular:SIGCOMM2013,dainotticcrpassive,Measuring-IPv6-Adoption}. 
Such assessment has become even more important recently, when the exhaustion of the readily available IPv4 address space puts
increased pressure on both ISPs and policy makers around the world.
The ISPs need to find open-ended ways to accommodate the needs for IPv4
connectivity of their customers, e.g., by increasing the utilization efficiency of their respective address blocks. The policy makers need to establish
regulatory guidelines for the emerging marketplace for IPv4 address
space~\cite{CCR-IPv4-scarcity}.

Recent studies that present Internet-wide statistics on IPv4 address space utilization have pushed the envelope in either measuring or estimating the total number of active IPv4 addresses~\cite{Capturing-Ghosts:IMC2014} and address blocks~\cite{dainotticcrpassive} in the Internet by relying on a number of diverse data sources.
However, the total number of active IPv4 addresses and blocks only partially captures address space utilization.
Moreover, with the exhaustion in allocation of IPv4 addresses, the situation will likely be changing and will reflect
the independent decisions of network operators, where 
the administration of the IP address space is under the control of the respective administrative domain (Autonomous System, of which about 51K can be found in the global routing table, currently). 
Varying resource demands and operational practices, as well as available supply of free and unused IP address space, blurs the notion of an ``active'' IP address. Today, we face a situation in which individual addresses and address ranges vary in their periods of activity and in that activity's nature and volume. For example, dynamic addressing, network reconfigurations, and users' schedules dramatically affect periods of activity.  When active, traffic characteristics and volumes range widely, from lightly used addresses and sparsely-populated blocks to individual addresses and blocks used by proxy gateways connecting thousands of devices to the Internet.

As a result, questions about the number of IP addresses active at a point of time, let alone their usage characteristics, are still difficult -- if not impossible -- to answer. It is even more difficult to comment on address space usage
characteristics over time or on how to even choose the right time granularity to observe such activity. The lack of detailed
measurement of address space activity is a major obstacle to our understanding of the current state of the IPv4 address space
exhaustion. Tracking and understanding address space utilization on a detailed level is both important for ISPs, who need to make business-critical decisions such as how to adapt their address assignment practices to this situation, as well as for regulators, who have to rely on estimations and predictions when introducing new policies which will ultimately affect what will be deployed in practice.
A detailed understanding of address activity also serves as foundation for security-critical systems that rely on
the notion of IP addresses, \eg for client reputation~\cite{IP-reputation:2009,DNS-reputation:2010}, as well as for systems that rely on active IP addresses
to perform measurements, \eg geolocation systems~\cite{constrained-geolocation:ToN,Octant,katz:2006} and network troubleshooting
systems~\cite{reverse-traceroute-nsdi2010,Tinocular:SIGCOMM2013,Fan:IMC2010}. 

In this study, we provide an unprecedented, detailed, and longitudinal view of IPv4 address space activity, as seen through the
lens of a large commercial CDN that serves almost 3 trillion requests per day.  This unique vantage point enables us to measure Internet-wide IPv4 address activity at the granularity of individual IP addresses, over a period that spans one entire year. Our study provides a number of insights on the state and growth of the Internet in the face of increasing IPv4 scarcity.

Our main contributions can be summarized as follows:
\begin{enumerate}[label=(\roman*),topsep=0.5ex,itemsep=0.5ex,partopsep=0.5ex,parsep=0.2ex,leftmargin=*,align=left]

\item We find that, after years of constant linear growth, the total number of active IPv4 addresses has  stagnated since 2014. Also, we find that state-of-the-art active measurement campaigns miss up to 40\% of the hosts that contact the CDN.

\item We show that, despite the stagnation in active IPv4 addresses in 2014, 
the \textit{set} of active  addresses is far from constant. In fact, over the 
course of a year, more than 25\% of the active IP address pool changes. Our 
analysis shows that most client networks contribute, with varying degrees, to 
this ``address churn'' and that this churn is hardly visible in the global 
routing table.

\item We identify a variety of address block activity patterns, and can attribute them to network restructuring, user behaviors, and, in particular, various address assignment practices. Based on our observations, we introduce metrics that allow us to quantify prevalent addressing practices at scale and comment on additional utilization potential within these -- already active -- address blocks.

\item We augment our address activity metrics with corresponding traffic volumes and relative host counts, which we derive from
\textit{HTTP User-Agent} samples, observing a trend of increasing traffic for addresses already heavily trafficked.
Combining our three key metrics of address activity, we then derive Internet-wide demographics of the active IPv4 address space and discuss 
implications that our study has towards enhancement of current operational and measurement practices.

\end{enumerate}

\section{Rethinking Address Activity}
\label{sec:rethinking}

\begin{figure}
    \centering
    \includegraphics[width=\columnwidth]{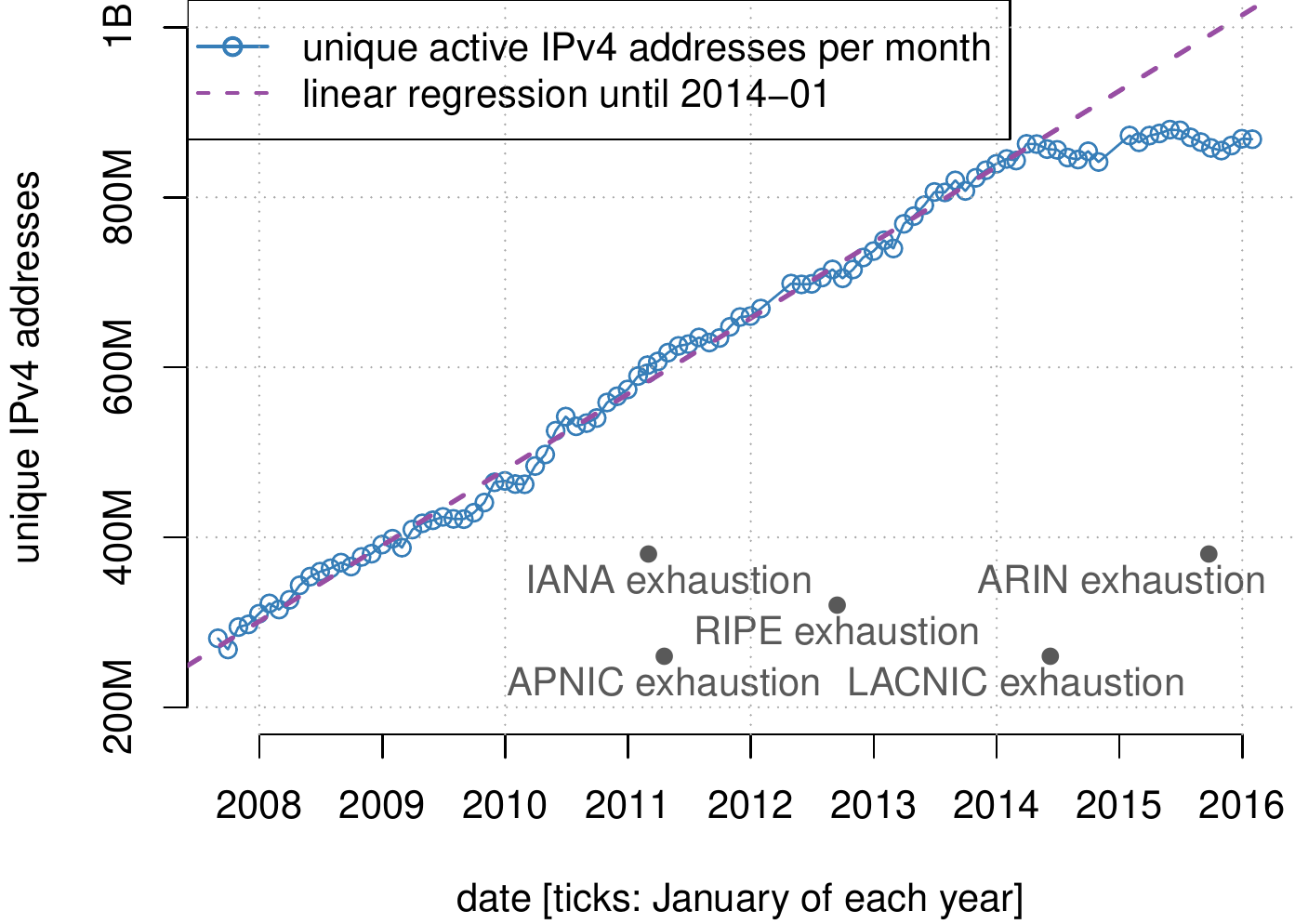}
    \caption{Unique active IPv4 addresses observed monthly by a large CDN.}
    \label{fig:active-IPv4-evolution}
\end{figure}

The study of the Internet's growth has attracted the interest of the research community since its early days. One fundamental dimension of this growth is the utilization of the available address space. As initially envisioned, every device on the Internet needs a globally unique IP address to be part of the Internet. Thus, the number of active addresses is a natural metric to track growth of the Internet. 
Figure~\ref{fig:active-IPv4-evolution} shows the number of monthly total active 
IPv4 addresses, as seen by a large commercial CDN.\footnote{Note:  The values 
in Figure~\ref{fig:active-IPv4-evolution} are about 5\% greater than those 
reported in Akamai's State of the Internet Report,~\cite{SOTI},
as the latter restricts to those addresses for which bandwidth is measured, which is also discussed in that report.
As the present work is not concerned with bandwidth, we omit this condition.}
For many years, an almost perfectly linear growth in terms of active IPv4 addresses was observed, conforming with our mental model
of a steadily growing number of used IPv4 addresses and corresponding address blocks. The most compelling observation from this
plot, however, is a sudden stagnation of the number of active addresses in 2014. This observation underlines a fundamental
point in the history of the Internet: The growth of active IPv4 addresses has subsided.

IPv4 address space scarcity has recently come to the full attention of the research and operations communities, as four out of the five \textit{Regional Internet Registries} (RIRs) that manage global IP address assignments have exhausted their available IPv4 address space~\cite{CCR-IPv4-scarcity}. Figure~\ref{fig:active-IPv4-evolution} is annotated with the respective exhaustion dates for each RIR. The prospect of exhaustion fueled intense discussions about how to ensure unhindered growth of the Internet by introducing technical as well as political measures to satisfy the ongoing demand, until we reach sufficient IPv6 adoption~\cite{Measuring-IPv6-Adoption}.\footnote{In this work, we exclusively focus on IPv4. We note that IPv6 address activity grew significantly during the year of 2015. The number of weekly active /64 IPv6 prefixes grew from 200M to more than 400M from September 2014 to September 2015. However, we emphasize that IPv6 /64 prefix counts are not directly comparable to IPv4 address counts. For more details on IPv6 client activity seen from the CDN, we refer readers to Plonka and Berger's work~\cite{plonka2015temporal}.}
A fundamental problem, however, is that getting an accurate and detailed picture of the current state of IPv4 address space
activity, and how this activity evolves over time, is quite difficult due to the Internet's decentralized structure. Past studies (which
we will discuss in detail in Section~\ref{sec:relwork}) typically relied on active or passive measurements to enumerate active
addresses and blocks. Given that we have now entered a period of stagnation, we argue that a sole enumeration of active IPv4 addresses does not draw an accurate picture of address space utilization and will be of little help, be it as a basis for policy decisions or for network operators making business-critical choices on how to manage their address space. 
We pose the following questions:

\vspace{0.1em}
\textbf{Q1} How effectively and at what granularity (Autonomous System, Prefix, 
/24 equivalent, IP address) can activity of the IPv4 address space be measured? 
Do geographic properties affect the visibility of IPv4 addresses?
(Section~\ref{sec:measuringaddressactivity})

\vspace{0.1em}
\textbf{Q2} At what timescales does activity manifest itself? What are the long-and short-term dynamics of IPv4 address space utilization? (Section~\ref{sec:macroscopicview})

\vspace{0.1em}
\textbf{Q3} Precisely, what operational practices contribute to these dynamics and which knobs could be adjusted to improve utilization? (Section~\ref{sec:microscopic})

\vspace{0.1em}
\textbf{Q4} What is the relationship between address space utilization, traffic volume, and the number of connected hosts? (Section~\ref{sec:traffic-and-UAs})

\vspace{0.1em}
\textbf{Q5} Can we extract meaningful address space demographics when combining our various metrics of address space activity? (Section~\ref{sec:clustering})

\section{Measuring Address Activity} 
\label{sec:measuringaddressactivity}

In this section, we first introduce the various methods that have been used in the past to measure and capture IP address space
activity. To this end, we discuss active and passive approaches used in related work. We then introduce our dataset, its
collection methodology and its advantages.  To assess the visibility of our dataset, we provide a comparison of our passive
IP address activity logs with active probing and provide a geographic breakdown 
of address visibility.

\subsection{Related Work}
\label{sec:relwork}

The most popular way of assessing IP address activity is by actively probing IP addresses (IPs), e.g., with ICMP queries. Heidemann et
al. presented a survey of IP address activity by systematically probing a subset of 1\% of the allocated IPv4 address space
with ICMP ping requests as early as 2008 \cite{Heidemann:2008:CSV:1452520.1452542}, which was followed by studies that
also capture aspects of network management, \eg diurnal activity patterns
\cite{Understanding-Block-Level-Address:SIGCOMM2010, quan2014internet} and Internet
reliability~\cite{Tinocular:SIGCOMM2013}. Recent improvements in active scanning techniques were introduced by Durumeric
et al. in ZMap~\cite{ZMap:USENIX2013}, that enable scanning of the entire IPv4 address space within
less than one hour or even in less than 5 minutes~\cite{adrian2014zippier}: a milestone in worldwide active measurement. 
Note that a reply from an IP address does not
necessarily indicate that a unique host is indeed active or even exists; tarpits~\cite{tarpits}, firewalls, and other middleboxes might send
replies to probe traffic destined to other IP addresses, or even entire IP address ranges. Also, active measurements cannot
capture activity at all timescales, as a reply might be dependent on many factors~\cite{schulman2011pingin,
quan2014internet}. It is also common that network administrators and home routers block ICMP traffic, thus, active measurements
are not always successful in detecting active address blocks~\cite{lostinspace}. Advanced active measurement techniques
that scan specific ports can be utilized to increase the detection success of active IPs~\cite{ZMap:USENIX2013}. 

Dainotti et al.~\cite{dainotticcrpassive,lostinspace} 
used passive measurements of packet captures and network flow summaries
recorded at various vantage points to infer IPv4 address space activity at the /24 level, finding 4.5M active /24 blocks from passive measurements in 2013. 
They detect and remove spoofed traffic, which can otherwise lead to overestimation of address activity.
Studying address activity only at the /24 level might be misleading, as the 
utilization within /24 blocks can vary widely.
To our best knowledge, only one related piece of work, by Zander et
al.~\cite{Capturing-Ghosts:IMC2014}, estimates the number of active IPv4 addresses (in contrast to address blocks). 
Combining seven different passively captured datasets and two active datasets, they use a statistical capture/recapture
model to account for invisible addresses and estimate the total number of active IPv4
addresses to be 1.2 billion as of 2014. %

A number of studies proposed techniques to identify dynamically assigned IPv4 addresses and uncover their dynamics.
Xie et al.~\cite{How-Dynamic-IP-Addresses:SIGCOMM2007} introduced a novel method, UDmap, that takes, as input,
user-login traces (e-mail logins in their study) and identifies the dynamic IPv4 addresses by associating the unique login
information of each user with the set of IPs it utilizes. They concluded that address dynamics exhibit a
large variation across networks, ranging from hours to several days. 
Jin et al.~\cite{identifying-dynamic-IP}
proposed and evaluated a technique 
based on distinct traffic activity patterns of static and dynamic addresses, 
in part when encountering outside scanning traffic. 
In a very recent
measurement study, Moura et al.~\cite{How-Dynamic-ISPs-Address:2015} proposed an active ICMP-based method to scan the addresses
of an ISP in search of blocks that rely on dynamic host configuration protocol (DHCP) to dynamically assign IPv4 addresses to
users and also to estimate DHCP churn rates. Padmanabhan et 
al.~\cite{IMC2016-Rama-DynIPs} used data gathered from RIPE Atlas probes to 
analyze the frequency and events associated with address assignment changes. 
They found that a number of ISPs around the world periodically reassign 
addresses after a fixed period, often a multiple of 24 hours. They also found 
that some address changes are correlated with network and power outages 
occurring at customer premises equipment. All the above mentioned works push 
the envelope in inferring dynamically assigned IP addresses, 
but either rely on user identification information, are an active measurement, 
or do not scale to the entire IPv4 address space.

Plonka and Berger count active World-Wide Web (WWW) client addresses by passive measurement and develop temporal and spatial
address classification methods~\cite{plonka2015temporal}. Their work has similarities to ours, here, in its use of CDN server logs (in
fact, the same logs we utilize) and in its spatio-temporal approach, but differs in that they study only IPv6 addresses.

\begin{table}[t]
\begin{center}
\resizebox{\columnwidth}{!}{
\begin{tabular}{p{3.8cm} | r r | r r | r r}
  & \multicolumn{2}{c}{\textbf{IP addresses}} & 
\multicolumn{2}{c}{\textbf{/24 blocks}} & 
\multicolumn{2}{c}{\textbf{ASes}} \\
\textbf{Description}& \textbf{total} & \textbf{avg.} & \textbf{total} & 
\textbf{avg.} & 
\textbf{total} & \textbf{avg.} \\
\toprule
\textit{Daily}: 08/17/15 - 12/06/15 & 975M & 655M & 5.9M & 5.1M & 50.7K & 
47.9K  \\
\textit{Weekly}: Jan - Dec 2015 & 1.2B & 790M & 6.5M & 5.3M & 53.3K & 47.8K  \\
\bottomrule
\end{tabular}
}
\end{center}
\caption{Datasets: Totals and averages per snapshot.}
\label{tab:dataset}
\end{table}

\subsection{The CDN as an Observatory}\label{sec:cdn-perspective}

The foundation for this study are server logs of one of the world's largest 
CDNs. In the year 2015, the CDN operated more than 200,000 servers in 120 
countries and 1,450 networks, serving content to end-users worldwide.
Each time a client fetches a Web object from a CDN edge server, a log entry is created, which is then processed and aggregated through a distributed data collection framework. After processing, we have access to the exact number of requests (``hits'') issued by each single IP address. In this work, we rely on two datasets, which are shown in Table \ref{tab:dataset}. For the year-long dataset, we have weekly aggregates of all IP addresses and 
for the daily dataset, we cover a period of 4 months. In the following, we refer to an IP address as \textit{active} if the CDN handled a request from that IP address in the given time interval. Correspondingly, we refer to an IP address as \textit{inactive} if there was not such a request.
Here, requests refer to successful WWW transactions, \ie an IP address will only be associated with a request if the client initiated a successful TCP and HTTP(S) connection and successfully fetched an object. Therefore, address activity is \textit{evident} from our log dataset and a major advantage compared to other passive measurements. The second advantage of our dataset is its \textit{granularity}, both space and time-wise. The logs contain numbers of requests on a per-IP level, illuminating a detailed picture of address activity.

\begin{figure}[!tbp]
  \subfigure[Visibility of IPv4 addresses, blocks, and networks.]{
    \includegraphics[width=0.9\linewidth]{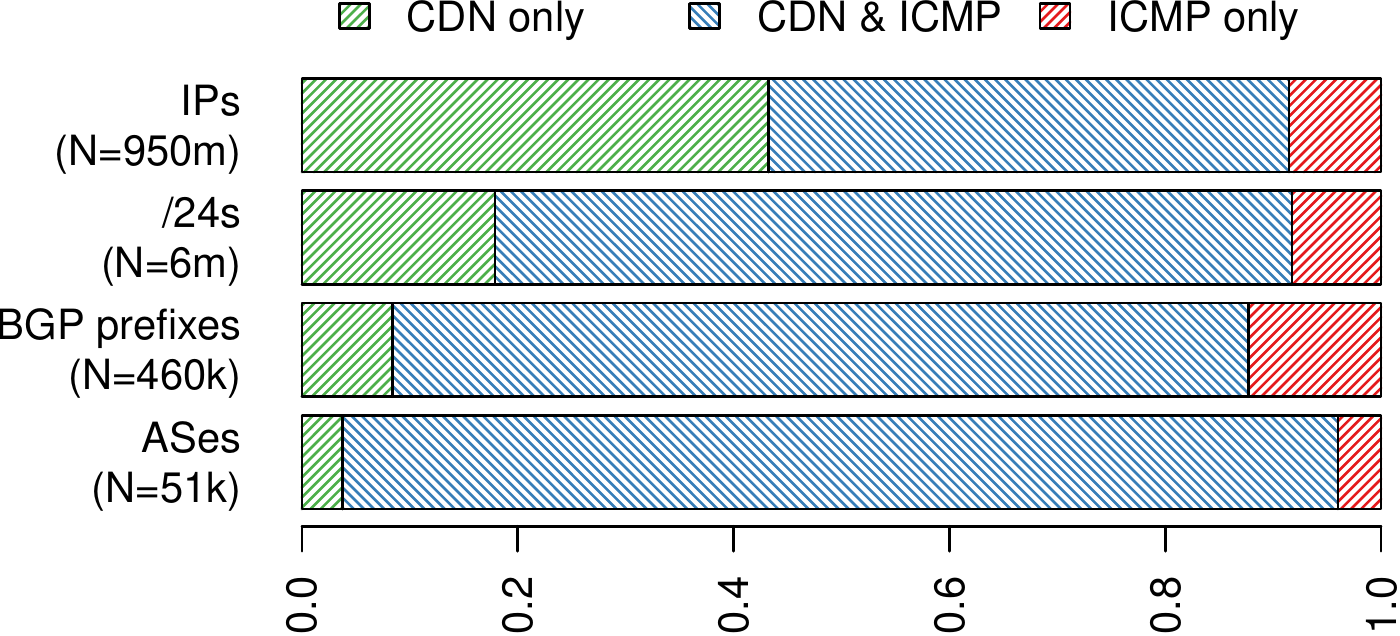}
    \label{fig:cdn_vs_zmap_bars}
  }
  \subfigure[Classification of IP addresses visible only in ICMP.]{
    \includegraphics[width=0.9\linewidth]{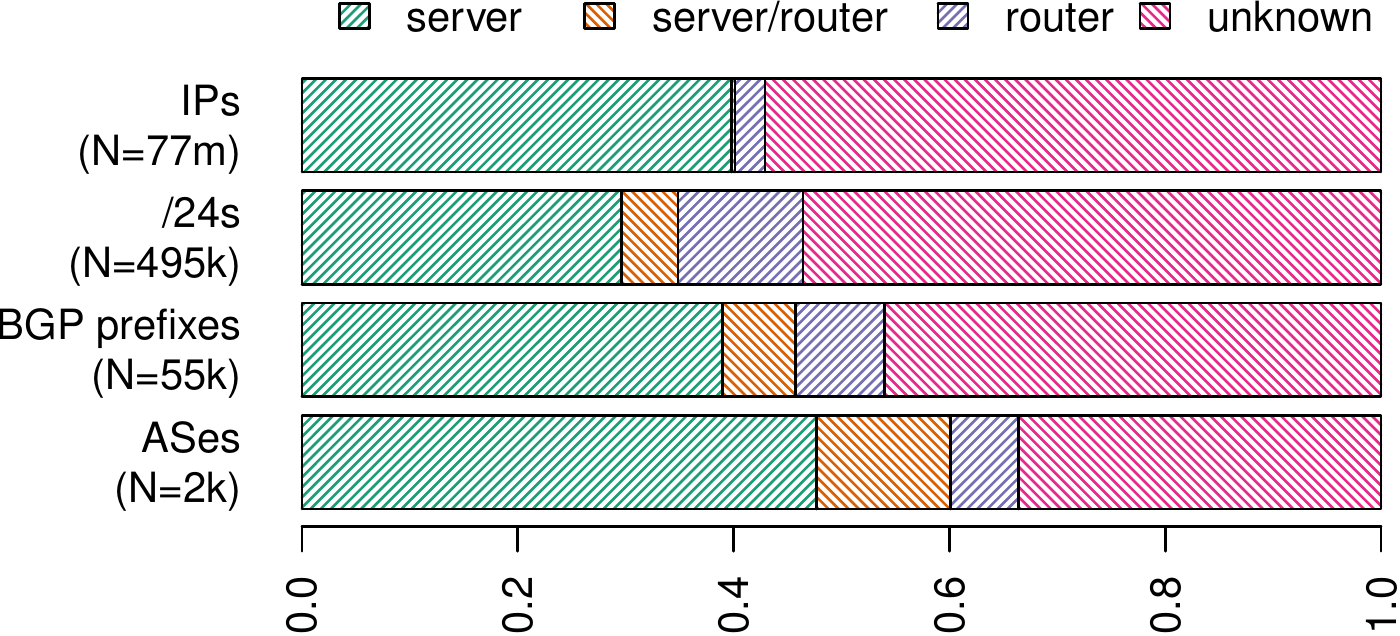}
    \label{fig:cdn_otheractivity}
  }
  \caption{Visibility into the IPv4 address space of the CDN compared with 
  active measurements (Oct. 2015).}
  \label{fig:CDN_vs_ZMap}
\end{figure}

To assess the view from our vantage point, next, we compare the set of addresses visible from the CDN to those which replies to ICMP queries. For this, we use the aggregated counts
of CDN-observed active IP addresses and compare them to the union of all IP addresses that were seen in 8 ICMP
scans, which we derive from the ZMap project \cite{ZMap:USENIX2013}.\footnote{We chose to show the comparison for October 2015 because the largest number of ICMP scans is available for this month.} 
Figure~\ref{fig:cdn_vs_zmap_bars} shows this comparison where the green bars are entities seen by CDN but not ZMap, the blue bars are entities seen by both the CDN and in ZMap, and the red bars are entities seen only by ZMap.
As illustrated in Figure~\ref{fig:cdn_vs_zmap_bars}, %
over 40\% of the 950 million IPv4 addresses show activity in the CDN logs but do not
appear to be active from ICMP probes. This difference is likely mainly attributable to hosts that sit
behind NAT gateways~\cite{IMC2016-Richter-NATs} and firewalls that do not permit replies to external requests via ICMP
or to hosts that respond only intermittently. 
While this pitfall is well-known
\cite{Fan:IMC2010,lostinspace}, we are not aware of any prior studies that quantify this effect at large scale. This incongruity is less pronounced when
aggregating the address space to /24 prefixes and ASes.\footnote{Here, we count a prefix/AS as active if we see activity from at least one IP address within the respective prefix/AS.} For routed prefixes and ASes, the number of (in)visible units is comparable for both methods, with ICMP outnumbering the CDN for the case of prefixes. Thus, measuring address space
activity on a per-prefix or even per-AS level, active measurements provide a 
significant coverage. On the per-IP level, however, active measurements are 
insufficient. We acknowledge that there is a bias in favor of the CDN logs with 
respect to WWW clients since we compare an entire month worth of CDN logs 
against 8 snapshots of ICMP scans, which will, naturally, not capture hosts that 
are active for only short periods of time. 

\subsection{Other Activity}

Despite the fact that much WWW content is hosted on the CDN platform and all the successful connections 
reported, our dataset has at least two limitations: 
\textit{(i)} the platform typically does not receive requests from Internet 
``infrastructure'' such as routers and servers
(though some routers and servers do obtain software updates from the WWW, 
and some servers obtain content from the WWW to, in turn, complete requests from their clients)
and \textit{(ii)} an IP may be assigned to a user who did not interact with the 
CDN platform.
To assess these, we next compare the portion of IPs that do reply to an ICMP request but are not present in
our CDN dataset, i.e., the red bars on the right in 
Figure~\ref{fig:cdn_vs_zmap_bars}. While this is roughly only
8\% of the IPv4 addresses in the combined CDN/ICMP dataset, we are interested in examining them further.

Figure~\ref{fig:cdn_otheractivity} shows a classification of these IP 
addresses, prefixes, and ASes. 
Here, we use additional data to identify servers and router
infrastructure. To identify servers, we rely on additional data from the ZMap project: IP addresses that replied
to server connection requests using HTTP(S), SMTP, IMAP(S) or POP3(S). To identify router IP addresses, we use one month
worth of the Ark~\cite{Ark} dataset and extracted all router IP addresses that appeared on any of the traceroutes (N=490M), \ie they replied with an ICMP TTL Exceeded error.
Close to half of the addresses that did not connect to the CDN, indeed, can be attributed to server or infrastructure
IP addresses. This fraction increases when aggregating to prefixes and ASes. We also note, however, that about half of these IP addresses 
did not show any server or infrastructure activity. These IP addresses might be 
(a) serving infrastructure that is not present in the Ark dataset, or infrastructure running 
other protocols than those probed by ZMap, or 
(b) practically unused IP addresses, or
(c) active IP addresses that simply do not connect to the CDN.

\subsection{Geographical View}

To gauge the geographic coverage of our dataset, as well as how it compares to 
active probing techniques regionally, we next dissect our dataset into 
geographic regions and countries. To accomplish this, we use allocation data 
provided by the RIRs \cite{nro_allocations} to assign regions and countries to 
each IP address.\footnote{We acknowledge that the exact geographic location of 
an IP address does not necessarily correspond to the country where the IP 
address was registered. We chose this dataset because it is publicly available, and we believe that this data is sufficient to 
highlight regional characteristics for the purpose of this study.} 
Figure~\ref{fig:cdn_vs_zmap_perrir} shows for each RIR the number of IP 
addresses that were visible both in the CDN dataset as well as responded to 
ICMP (bottom red bars), the number of IP addresses that were only visible in 
the CDN logs (middle blue bars), as well as those IP addresses that were 
invisible in our logs, but appeared in ICMP scanning campaigns (green bars on 
top). We observe that the CDN logs provide substantial additional visibility in 
all regions. When put in relation to the total number of active IP addresses 
per region, this effect is particularly pronounced in the African region, where 
the CDN logs increase the number of visible active IP addresses by more 
than 150\%.

\begin{figure}[!tbp]
  \subfigure[Visibility of IPv4 addresses grouped by RIR.]{
    \includegraphics[width=0.9\linewidth]{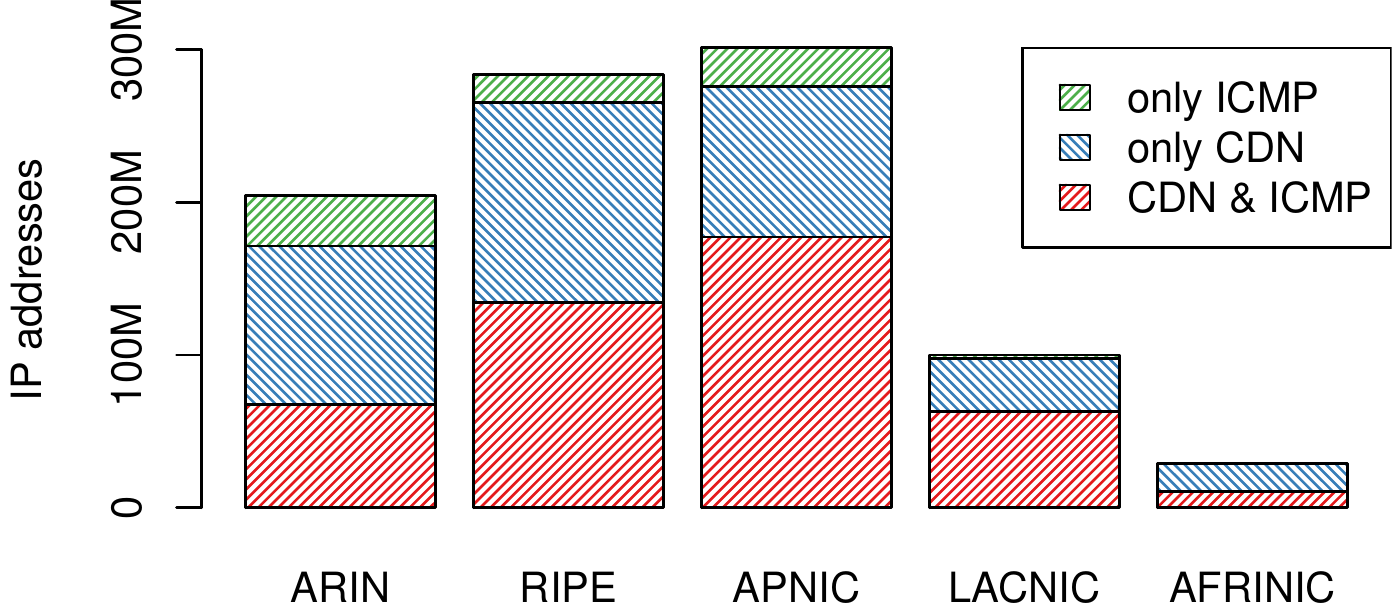}
    \label{fig:cdn_vs_zmap_perrir}
  }
  \subfigure[Top countries, annotated with their rank of broadband and cellular subscribers (ITU, 2015).]{
    \includegraphics[width=0.9\linewidth]{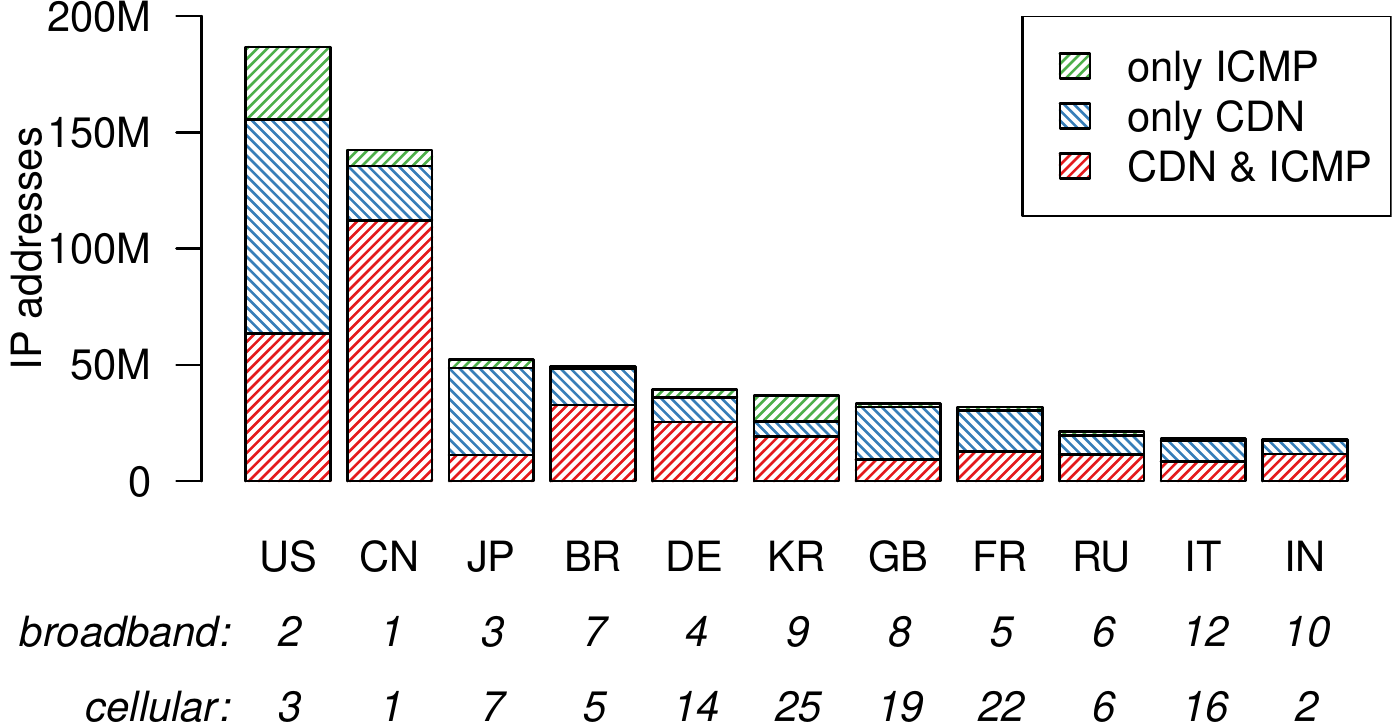}
    \label{fig:cdn_vs_zmap_percountry}
  }
  \caption{IP address activity by geographic region.}
  \label{fig:CDN_vs_ZMap_Geo}
\end{figure}

In Figure~\ref{fig:cdn_vs_zmap_percountry}, we show the partition of addresses for the top countries in terms of the number of addresses seen in the CDN logs and the ICMP scans. In addition, 
we annotate each country with its rank of fixed broadband and cellular 
subscribers, based on ITU data \cite{ITU_data}. Here, we see that the top 
countries ranked by broadband subscribers are also the top countries visible in 
the CDN logs. Thus, the coverage of our dataset largely agrees with ITU 
estimates on global Internet subscribers. This effect is much less-pronounced 
when ranking countries per cellular subscribers, perhaps because most cellular 
networks deploy Carrier-Grade NAT \cite{IMC2016-Richter-NATs}, 
blurring the relationship of subscribers to IP addresses. Secondly, we also see 
that the fraction of ICMP-responding IP addresses varies heavily per country. 
In China, for example, we find that close to 80\% of the IP addresses do 
respond to ICMP requests, whereas in Japan only about 25\% of the IP 
addresses reply to ICMP requests. An observation to keep in mind when, e.g., 
using active measurement techniques to reason about Internet penetration or 
address space utilization in specific parts of the world.

\section{Macroscopic View of Activity}
\label{sec:macroscopicview}

To bootstrap our analysis, we study IPv4 address activity on a broad scale in this section. In particular we focus on how many addresses our vantage point observes as well as how consistent the set of active IPs is over time. Then, we focus on spatial properties of the observed dynamics and compare our observations with what is visible from the global routing table. 

\begin{figure*}
  \subfigure[Daily active IPv4 addresses and up/down events.]{
    \includegraphics[width=0.315\linewidth]{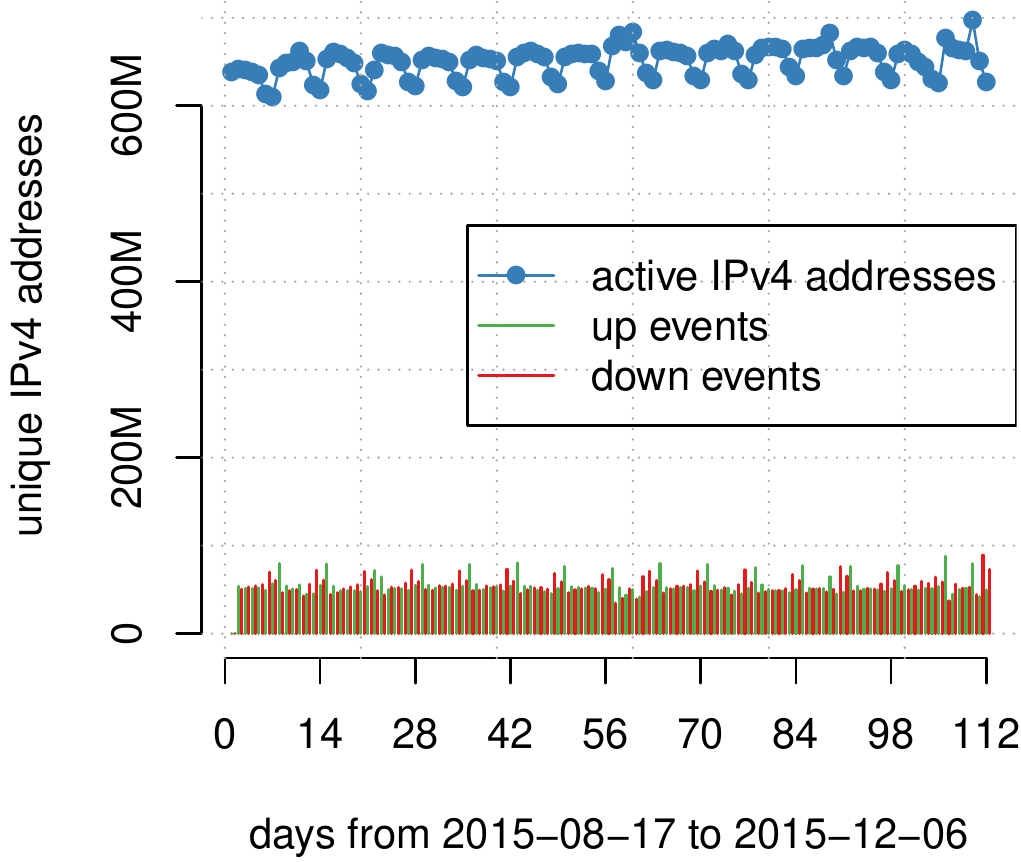}
    \label{fig:daily_ipv4}
  }
  \hfill
  \subfigure[Median up/down events between subsequent snapshots for different aggregation windows (union of active IPs within window).]{
    \includegraphics[width=0.315\linewidth]{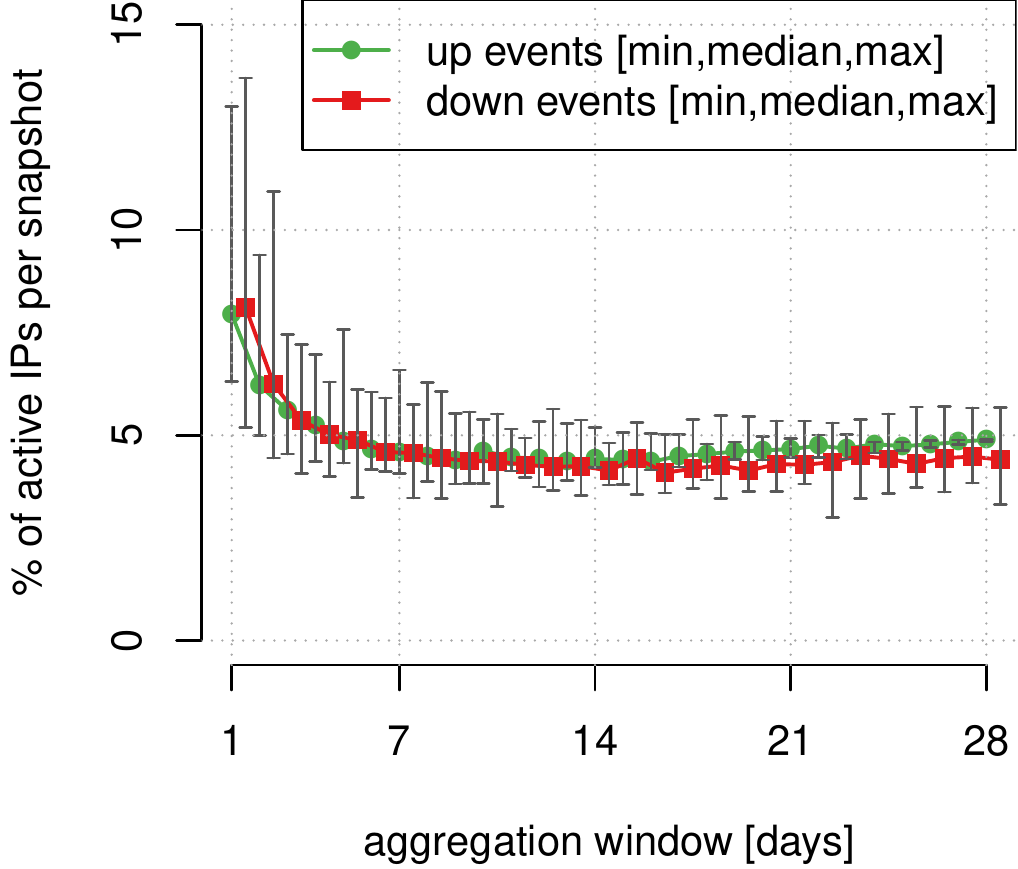}
    \label{fig:up_down_events_allscales}
  }
  \hfill
  \subfigure[Difference in active IPv4 addresses compared to first snapshot of our period (weekly).]{
    \includegraphics[width=0.315\linewidth]{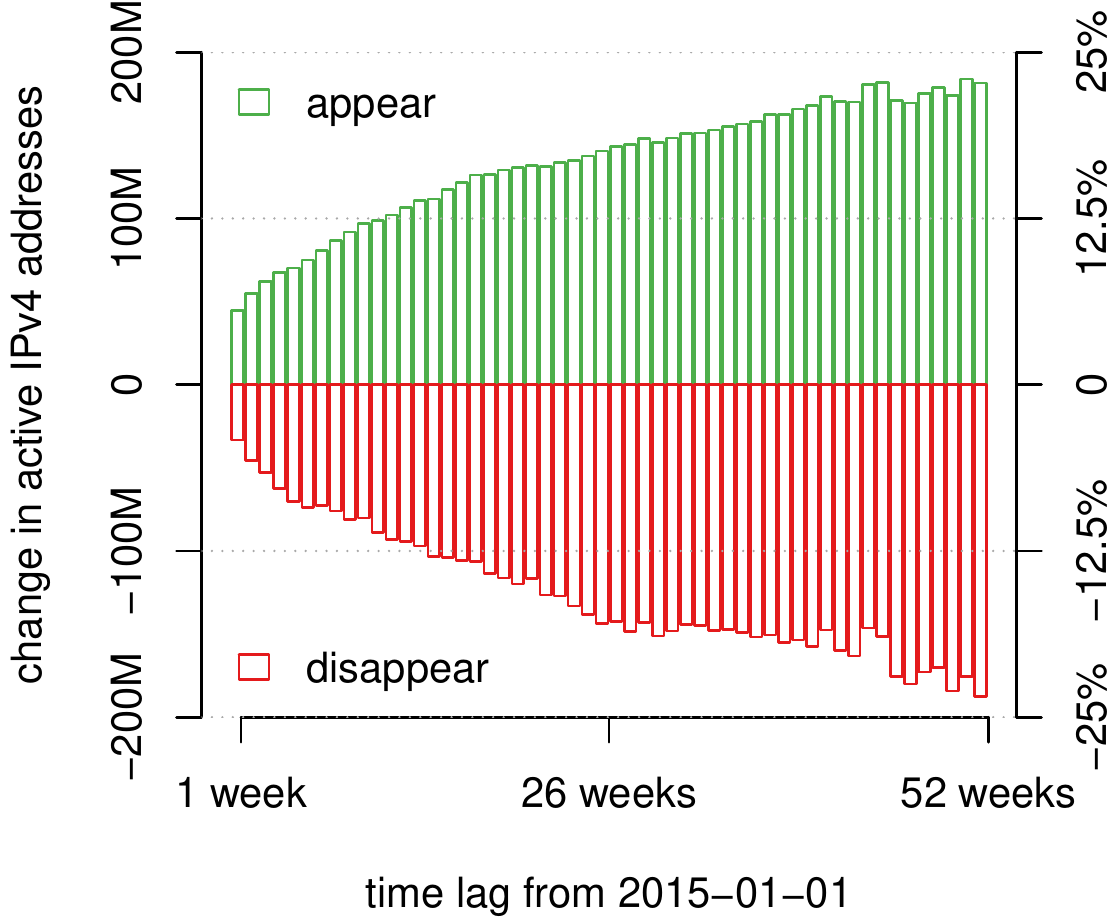}
    \label{fig:long_term_diff}
  }
  \hfill
\caption{Activity and churn in active IPv4 addresses.}
\end{figure*}

\subsection{Volatility of Address Activity}

To assess address activity over time, we show in 
Figure~\ref{fig:daily_ipv4} the daily number of unique IPv4 addresses that 
contact the 
CDN over the course of 16 weeks. We observe about 650M unique active IPv4 
addresses on a daily basis, and less on weekend days. 
Although Figure~\ref{fig:daily_ipv4} shows a relatively constant 
\textit{number} of active IPv4 addresses, the \textit{set} of addresses can 
vary. To capture changes in the population of active addresses, we define an 
\textit{up event} if an address is not seen in a given window of time, e.g., a 
day or 7 days, but then is seen in the subsequent window. Likewise a 
\textit{down event} occurs if an address is seen in a given window of time, 
but not seen in a subsequent window. Figure~\ref{fig:daily_ipv4} shows an 
average of 55M daily up events, likewise for down events. %
Hence, each day we see 55M addresses showing activity that were not active 
the day before. Another 55M addresses are active that day, but not on the next day.

We next assess whether this churn appears only on short timescales (\ie due to 
short-term inactivity of certain IP addresses) and disappears on longer 
timescales, e.g., when comparing subsequent weeks to each other as opposed to 
days. In Figure~\ref{fig:up_down_events_allscales}, we partition the $112$ days 
of Figure~\ref{fig:daily_ipv4} into non-overlapping windows, of a given size. 
For a window size of $7$ days, for example, there would be $16$ windows, or 
snapshots.  In each window, we note the union of all active IP addresses.  Then,
for window $i$ and $i+1$ we compute the percentage of addresses that had an up 
event as 100 $\times$ (the number of addresses in window $i+1$ that are not present in 
window $i$) divided by the number of addresses in window $i+1$. Hence, for a 
window size of $7$ days, we obtain $15$ such percentages. 
We then note the minimum, median, and maximum of these percentages. We do the 
analogous computation for down events. In 
Figure~\ref{fig:up_down_events_allscales}, the two red and green points at $x=1$ 
on the 
x-axis show 
the min, median, and max of the percentage of addresses that had up/down events 
on a daily basis, corresponding to Figure~\ref{fig:daily_ipv4}. On an average 
day, about 8\% of the active addresses ``come,'' another 8\% ``go.''
We see that the maximum values for up/down events are as high as 14\%, reflecting changes from weekdays to weekends and vice versa.
The red/green dots at $x=7$ show these statistics when we aggregate our dataset 
into weeks and compare subsequent weeks. The stunning observation from this 
figure is that, while churn is more apparent on short timescales (particularly 
for window sizes for 1 and 2 
days, related to day-of-the-week effects), the dynamics in up/down 
events do not decay to zero for higher aggregates. Indeed, we observe that the 
churn level for aggregates larger than 7 days remains constant at roughly 5\%. 
Thus, whichever aggregation level we choose (days, weeks, months), the set of 
active IP addresses is in constant change, both on short, as well as on long 
time scales.

To highlight the long-term effects, Figure~\ref{fig:long_term_diff} shows,
weekly,
the number of newly appearing and disappearing IP addresses as compared 
to the first week of 2015. That is, for each week in 2015 (x-axis), we show
the number of addresses that were \textit{not} active in the first week 
(positive y-axis, \textit{appear}), but in the given week and also the number 
of IP addresses that were active in the first week, but \textit{not} in the 
given week (negative y-axis, \textit{disappear}). In fact, the set of active 
addresses has changed by as much as 25\% over the course of 2015.

\subsection{Dissecting Address Volatility}
\label{sec:dissectingvolatility}

\begin{figure*}
  \subfigure[CDF: Median \% of up events per AS and snapshot (only ASes with $>$ 1000 active IPs, N = 8.6K).]{
    \includegraphics[width=0.315\linewidth]{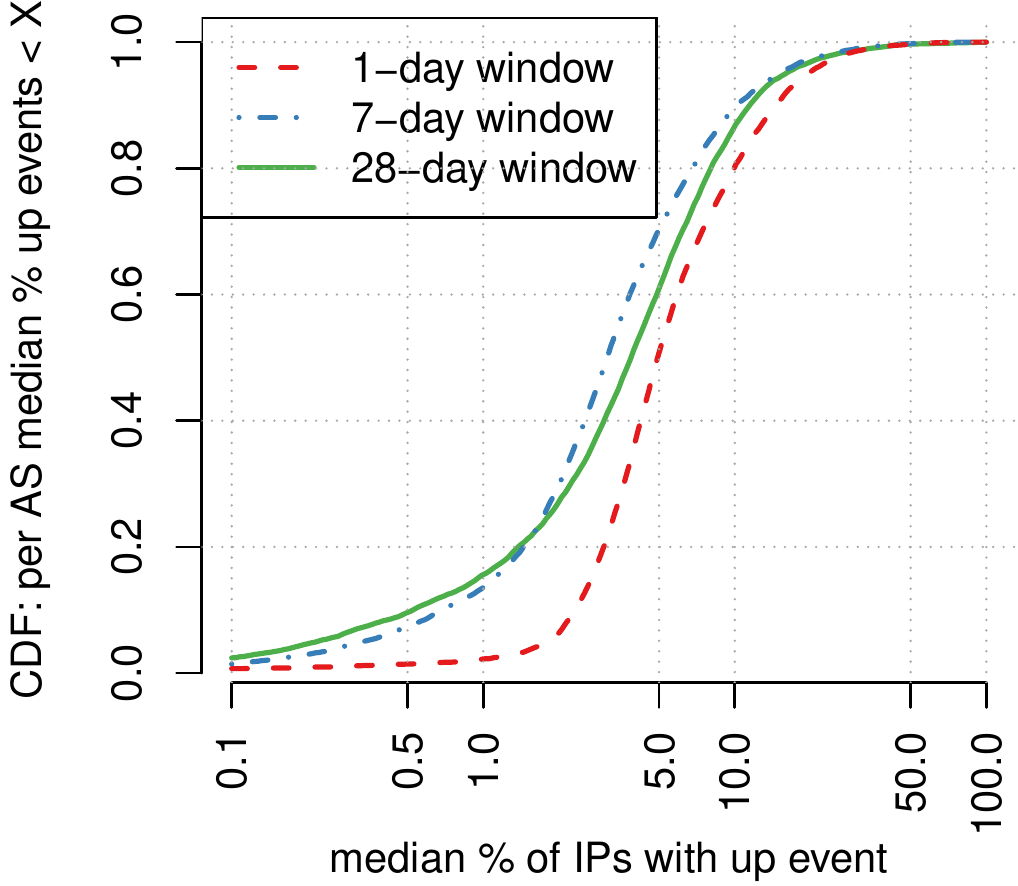}
    \label{fig:churn_per_as}
  }
  \hfill
  \subfigure[Size distribution of up events for different time ranges.]{
    \includegraphics[width=0.315\linewidth]{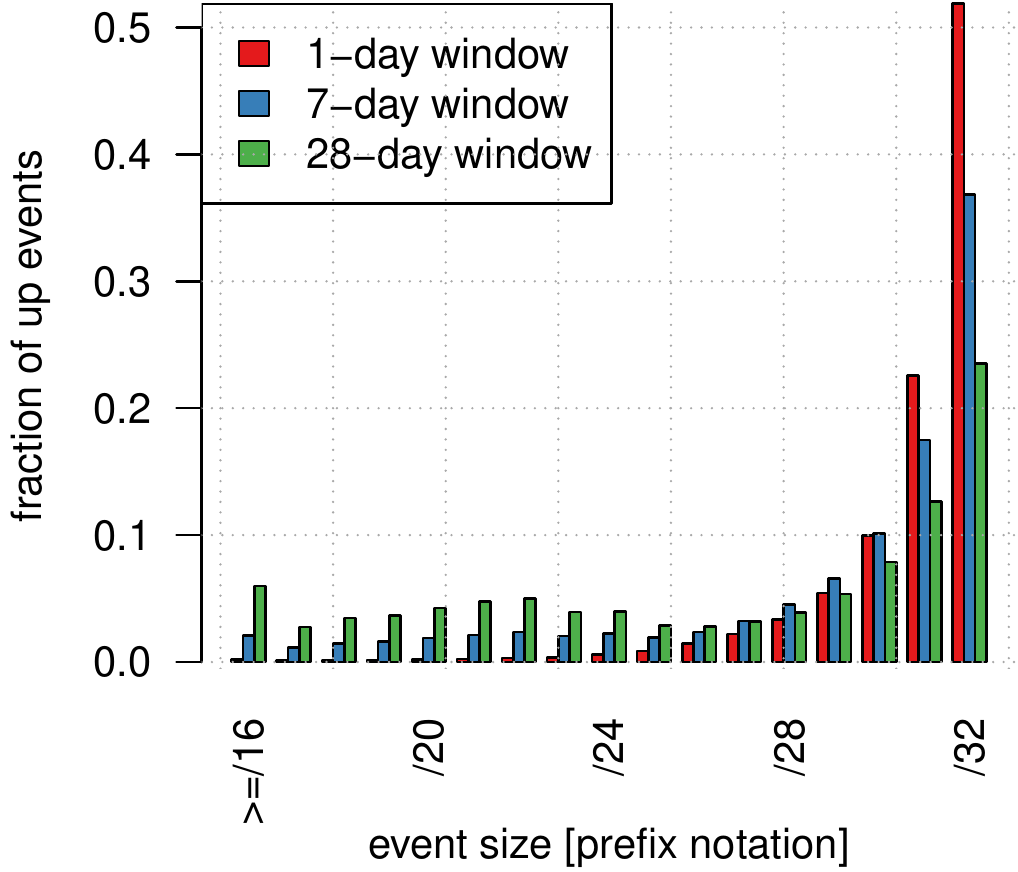} %
    \label{fig:churn_eventsize}
  }
  \hfill
  \subfigure[\% of up/down events that go together with a change in BGP for different aggregation windows.]{
    \includegraphics[width=0.315\linewidth]{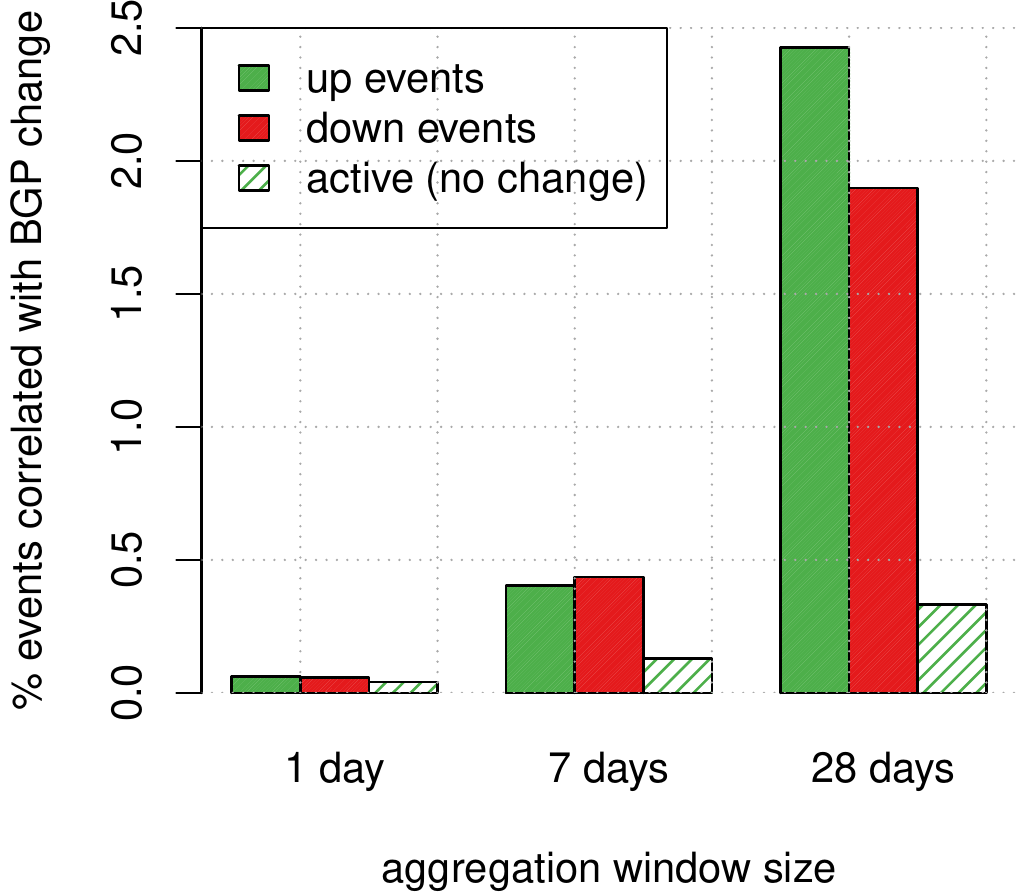}
    \label{fig:churn_vs_bgp}
  }
\label{fig:churn-observations}
\caption{Address churn properties.}
\end{figure*}

Having seen that the active portion of the IPv4 address space is highly volatile in nature, we next study some macroscopic features of the observed dynamics. In particular, we study \textit{(i)} if networks contribute similar levels of churn, \textit{(ii)} the size of up/down events, in terms of prefixes and \textit{(iii)} if this churn is also reflected in the global routing table.

\parax{A network view of churn} In Figure~\ref{fig:churn_per_as}, per
Autonomous System (AS), we show the median percentage of IP addresses with an 
\textit{up} event for each snapshot. That is, we partition the set of addresses 
into ASes, and we repeat the calculation of 
Figure~\ref{fig:up_down_events_allscales} for addresses in each AS, and obtain 
a median percentage for each AS.  Figure~\ref{fig:churn_per_as} shows the CDF 
of these medians. We only consider ASes for which we saw at least 1K active IP 
addresses during our observation period and we only show \textit{up} events, 
the CDF for \textit{down} events is similar.
The takeaway from this figure is that highly dynamic IP address activity is 
not a phenomenon limited to a small number of ASes - rather, about 10\% to 20\% 
(depending on window size) of the ASes have a 10\% or higher median percentage 
of IPs with an up event. About half of the ASes have a churn rate below 5\%, 
the other half a higher one. We observe this for different aggregation windows, 
with a slight decrease in volatility for some ASes on higher aggregation 
levels. Thus, churn is a ubiquitous phenomenon, which we observe for a large 
number of networks.

\parax{A prefix view of churn} 
So far, we have considered up and down events on a per-address basis (for 
different time window sizes).  Next, we are interested in whether up and down 
events really only affect single addresses, or rather entire address ranges. In 
particular, we are interested in entire prefixes that have been inactive and 
then some or all of the addresses become active, which we expect would likely 
indicate network operator actions as opposed to independent, individual user 
behavior.

To accomplish this, for each per-address up event, we find the smallest prefix 
mask $m$ (where a smaller mask corresponds to a prefix that contains more 
addresses) in which all addresses either had an up event or showed no activity 
in both snapshots. 
Figure~\ref{fig:churn_eventsize} shows a histogram of the fraction of 
per-address up events, for a given window size, where we assign each up event 
to its tagged prefix mask $m$ (the histogram for down events looks similar). 
For example, for a window size of 1 day, more than 70\% of the per-address up 
events are associated only with a mask $\geq$ /31, indicating that these 
dynamics typically only affect individual IP addresses.

For larger aggregates (e.g., 28-days), we still see more than a third of the up 
events in the $\geq$ /31 range, however we also observe some up events spanning 
larger ranges of addresses, with more than 38\% of month-to-month up events 
affecting larger address blocks with a mask $\leq$ /24.
Thus, a key observation when studying churn across different time aggregates is
that a significant proportion of long-term events (38\% on a month-to-month 
aggregation) affect entire prefix masks $\leq$ /24, some of them as large as an 
entire /16 prefix. These ``bulky'' events hint towards 
changes in address assignment practice (e.g., network restructurings), as 
opposed to churn caused by individual ON/OFF activity of a single IP address. 
While this is an expected property and holds for some portion of the 
month-to-month churn, we also notice that this certainly does not hold for all 
events on larger timescales. In fact, even on a month-to-month scale, more than 
36\% of the events only affect prefixes of size /31 or even /32, i.e., single IP 
addresses.

\parax{A routing table view of churn} Given that the active IP address 
population changes by about 25\% over the course of a year (per 
Figure~\ref{fig:long_term_diff}), an appropriate question is whether these 
dynamics 
are also reflected in the global Border Gateway Protocol (BGP) routing table. 
To assess this, we next 
associate each IP address with 
its origin AS using daily snapshots of the global routing table.\footnote{We 
rely on daily snapshots from a RouteViews collector in AS6539. For larger 
window sizes, we determine the origin AS for a given IP address using a 
majority vote of all contained daily IP-to-AS mappings.} We show in 
Figure~\ref{fig:churn_vs_bgp} the fraction of up/down events that go together 
with a 
BGP change. Here, we consider both route announcements, withdrawals, as well 
as origin AS changes as a ``BGP change'' event. The green bars show the 
percentage of \textit{up} events that go together with a BGP change, and the 
red bars show the percentage of \textit{down} events. In addition, we also plot 
the fraction of steadily active (no up/down event) IP addresses and for what 
fraction of them we observe changes in the routing table. While we can clearly 
see that \textit{(i)} IP addresses with up/down events are much more likely to 
correlate 
with events in 
the routing table when compared to steadily active addresses and \textit{(ii)} 
that, on higher aggregation levels, up/down events are more likely to correlate 
with BGP changes, reflecting network changes, we find that \textit{(iii)} only a 
tiny minority of these events are visible in the global routing table (less 
than 2.5\% for monthly aggregation levels). Thus, the vast majority of 
volatility in IP address activity is entirely hidden from the global routing 
table.

\begin{table}
\begin{center}
\resizebox{\columnwidth}{!}{
\begin{tabular}{p{4.5cm} | r r}
\textbf{}  & \textbf{appear} & \textbf{disappear} \\
\toprule
\textbf{total}  & 139M & 129M \\
\textbf{entire /24 prefix affected}  & 65\% & 54\% \\
\toprule
\textbf{BGP no change}  & 87.1\% & 90.4\% \\
\textbf{BGP origin change}  & 3.3\% & 7.1\% \\
\textbf{BGP announce/withdraw}  & 9.6\% & 2.5\% \\
\bottomrule
\end{tabular}
}
\end{center}
\caption{IP addresses that appeared/disappeared comparing Jan/Feb 2015 and 
Nov/Dec 2015, percentage of those IP addresses where the entire containing /24 
prefix appeared/disappeared, and corresponding BGP changes.}
\label{tab:longtermchurn}
\end{table}

\subsection{Volatility During One Year}

Next, we study those IP addresses, in particular, that were first inactive for a long period and then became active, as well as IP addresses that showed activity but then went inactive. For this, we pick the first two months of our observation period (January, February 2015) and the last two months of our observation period (November, December 2015), where we take the union of all active IP addresses that were seen within each snapshot. We then compare the two snapshots, and also the associated BGP activity.
Table \ref{tab:longtermchurn} summarizes our results. 
Continuing the trend shown in Figure~\ref{fig:churn_eventsize}, that churn becomes bulkier on longer time scales,
we observe that more than
half of the events (65\% and 54\%, respectively) affected entire address blocks, 
and are, thus, more likely to be caused by operational changes. However, another 
large chunk of long-term volatility affects smaller aggregates, down to single 
IP addresses. The main result in Figure~\ref{fig:churn_vs_bgp}, that 
only a small minority of these events coincide with BGP changes, also pertains 
at the year-long time scale in Table \ref{tab:longtermchurn}. In fact, most of 
these IP addresses were -- and are -- still routed by the same AS.

More than 30K ASes announce IP addresses that show long-term volatility in our 
dataset without any change in BGP configuration.
The top 10 ASes, in terms of IP addresses of the class that appear or disappear,
contribute about 30\% of the total addresses in each class.  These top 10 ASes 
include major ISPs connecting both residential and cellular mobile users. In 
fact, we find that ASes contributing the most IP addresses to the appear class 
are also those ASes contributing the most addresses to the disappear class. 
Focusing on our two sets of top 10 ASes, we note that we find 7 of those 
contributing to the appear class are also among the top 10 contributing to the 
disappear class. 
Thus, while contributing large number of IP addresses with 
high volatility, the total number of active IP addresses for these ASes varied 
only marginally, in the order of a few percentages. Hence, we can attribute the 
majority of long-term volatility to AS-internal dynamics, as opposed to 
networks entering the market or going out of business.

\section{Microscopic View of Activity}
\label{sec:microscopic}

\begin{figure*}[t]
\center
  \subfigure[Statically assigned address block (German University, FD=29, STU=0.04).]{
    \includegraphics[width=0.22\linewidth]{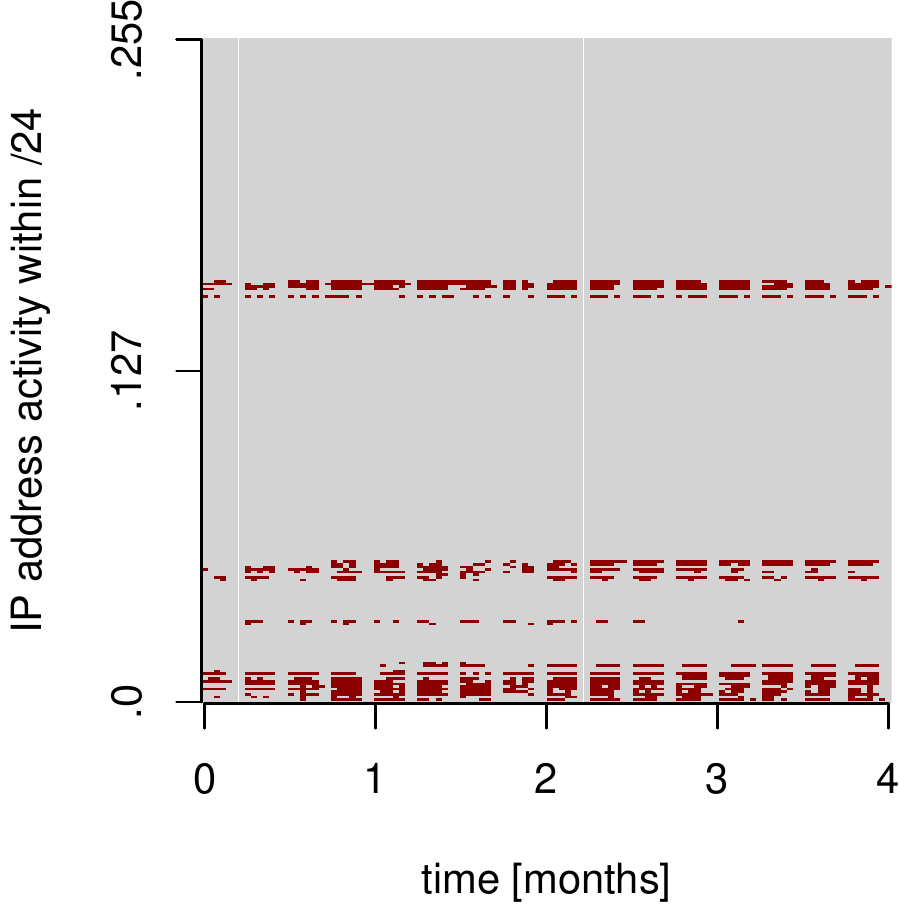}
    \label{fig:static-light} %
  }
  \hfill
  \subfigure[Dynamically assigned address block (US University, FD=254, STU=0.18).]{
    \includegraphics[width=0.22\linewidth]{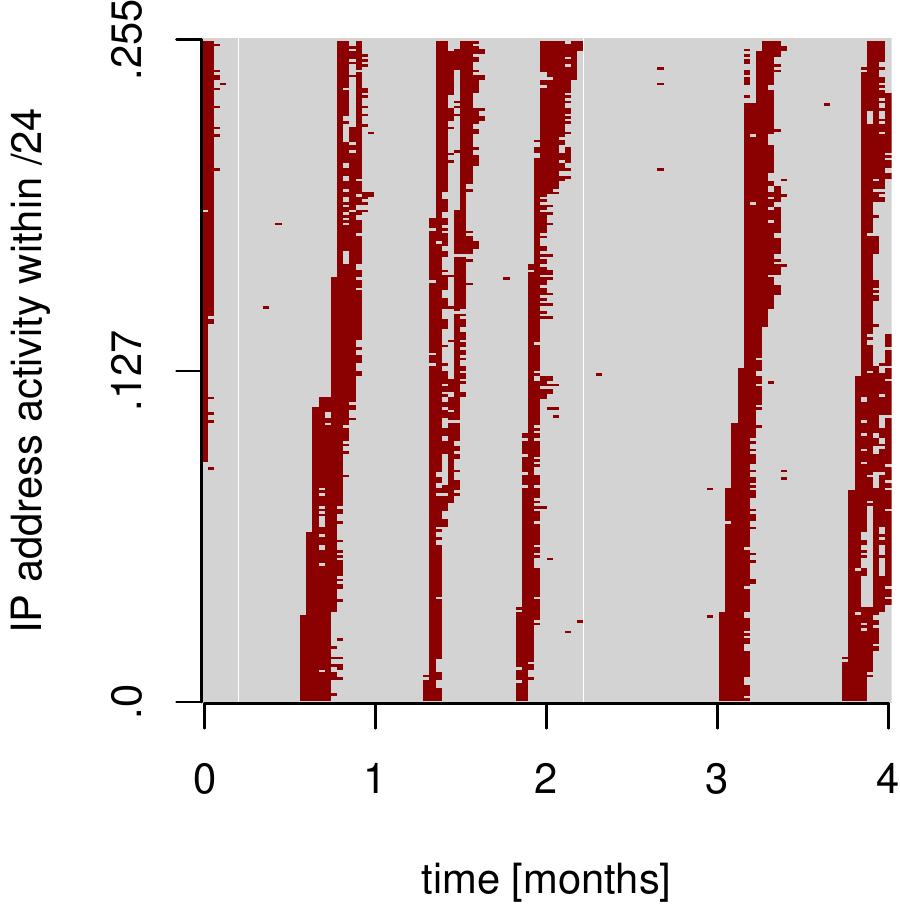}
    \label{fig:DHCP} %
  }
  \hfill
  \subfigure[Dynamically assigned address block with residential users (US ISP, FD=175, STU=0.26).]{
    \includegraphics[width=0.22\linewidth]{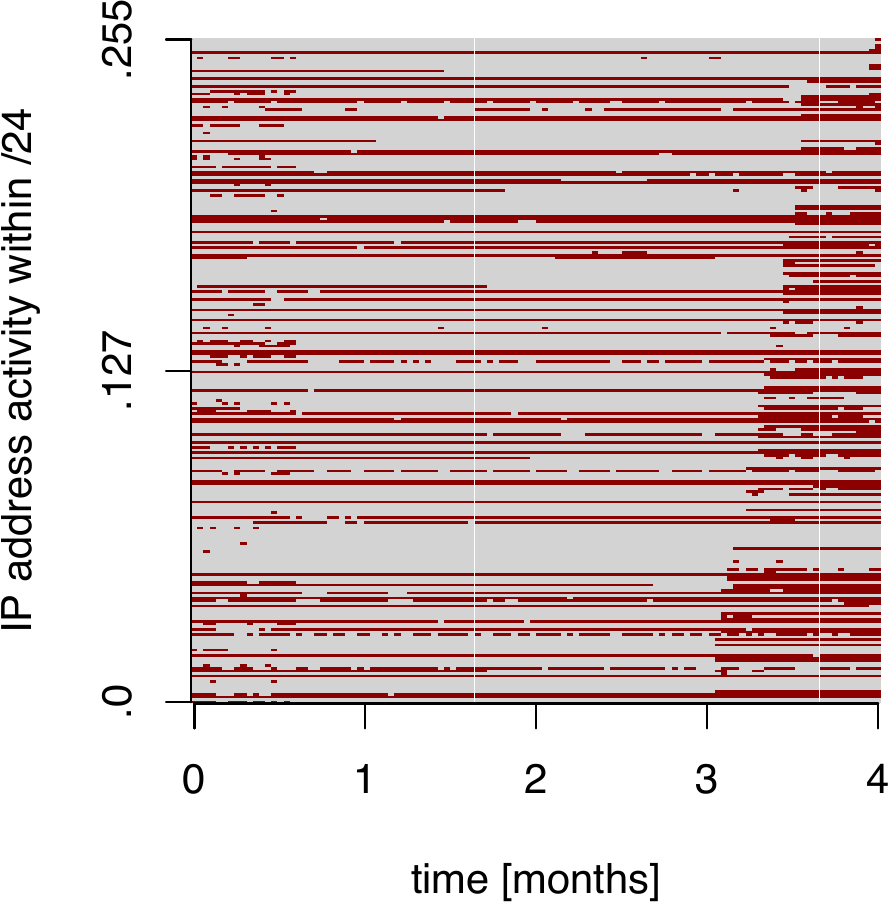}
    \label{fig:churn_usisp} %
  }
  \hfill
   \subfigure[Dynamically assigned address block with residential users (German ISP, FD=254, STU=0.75)]{
    \includegraphics[width=0.22\linewidth]{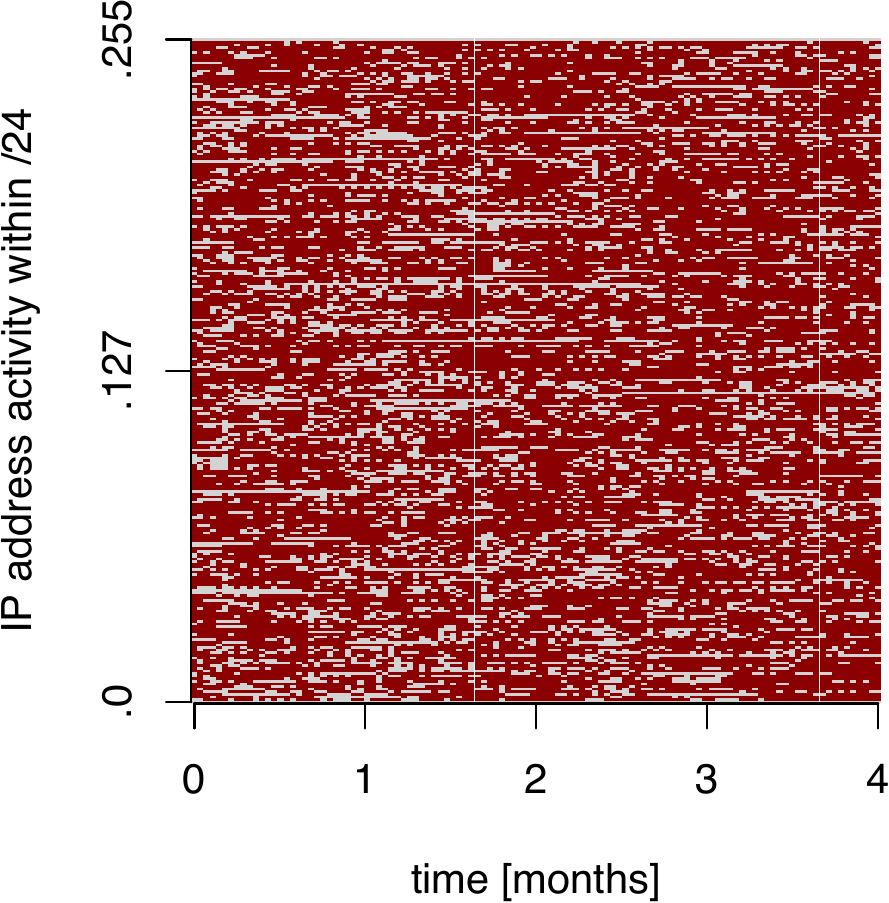}
    \label{fig:dynamic-heavy} %
  }
\caption{Regular activity patterns: Interplay between address assignment practice and user-behavior.}
\label{fig:microscopic-view}
\end{figure*}

Given observations of churn in the active IPv4 address space, we now drill down into their root causes. The way that IP addresses are allocated and used across network operators is not uniform. There are many factors that contribute to how a network operator assigns IPs to client hosts, \eg address pool size, client population, type of clients (enterprise or residential), privacy considerations (residential users may have their address lease expire after a maximum duration), or other operational practices (static/dynamic address assignment). Thus, it is challenging to characterize the IP assignment strategies within a single network, let alone an entire address space.

To offer a glimpse of how activity is typically manifest in different areas of 
the IPv4 address space, in Figure~\ref{fig:microscopic-view}, we show examples 
of activity patterns in four address blocks. Here, we examine specific /24 
prefixes, as it allows us to present a spatio-temporal view of activity in 
address-level detail. To generate these plots, we rely on our 4 months worth of 
daily IP address activity (x-axis). We then align all IP addresses within the 
selected /24 on the y-axis in increasing order. Having this ``activity matrix'' 
in place, we plot a red point for each day on which a given address was active. 
With these examples in mind, we introduce two root causes for churn in address 
activity:

\para{Regular activity patterns: Address assignment practice.} The four 
examples in Figure~\ref{fig:microscopic-view} show striking differences in 
daily address activity. While we see a non-uniform, light utilization in 
Figure~\ref{fig:static-light}, with a day-of-week pattern for few active 
addresses, we see heavier utilization in 
Figures~\ref{fig:DHCP},~\ref{fig:churn_usisp}, and~\ref{fig:dynamic-heavy}, 
with a variety of activity patterns involving dynamic assignment from address 
pools. While Figure~\ref{fig:DHCP} shows a round-robin IP address assignment in 
an underutilized pool, Figure~\ref{fig:churn_usisp} shows dynamic addressing 
with a very long lease time (i.e., the duration for which a specific subscriber 
holds an IP address), with some IP addresses having almost continuous activity 
and others having infrequent activity. Figure~\ref{fig:dynamic-heavy} shows 
another mode of dynamic addressing, wherein the ISPs lease time is set to a 
maximum of 24 hours, thus causing hosts to be frequently reassigned a different 
IP address. We refer to the activity patterns in 
Figure~\ref{fig:microscopic-view} as \textit{in situ} activity, as they result 
from address assignment practice and its interplay with end-user behavior in 
one administratively-configured {\em situation;} that is, we have no evidence 
that the situation, nor the activity pattern, changed due to network 
reconfiguration.
A key observation is that \textit{in situ} activity in address blocks varies significantly amongst those that have different address assignment configurations.

\para{Changed patterns: Modification of assignment practice.} As shown in 
Figure~\ref{fig:microscopic-view-admin-change}, we also observe activity 
patterns that are temporally or spatially inconsistent.  This is some evidence 
that the patterns' dynamics are not the result of constant address assignment 
policy, but, rather, are the result of address (a) reallocation, (b) assignment 
reconfiguration, and/or (c) repurposing.

We next study address activity pattern at large scale. In particular, we are first interested in detecting, which portions of the address space show a consistent address assignment pattern as opposed to blocks that show major changes in their activity pattern. We then dive into the former, activity patterns that are the result of address assignment practice in conjunction with end-user behavior. Here, we put a particular emphasis on the resulting \textit{utilization} of address blocks.

\begin{figure}[t]
\center
  \subfigure[German University, FD=256, STU=0.32.]{
    \includegraphics[width=.47\linewidth]{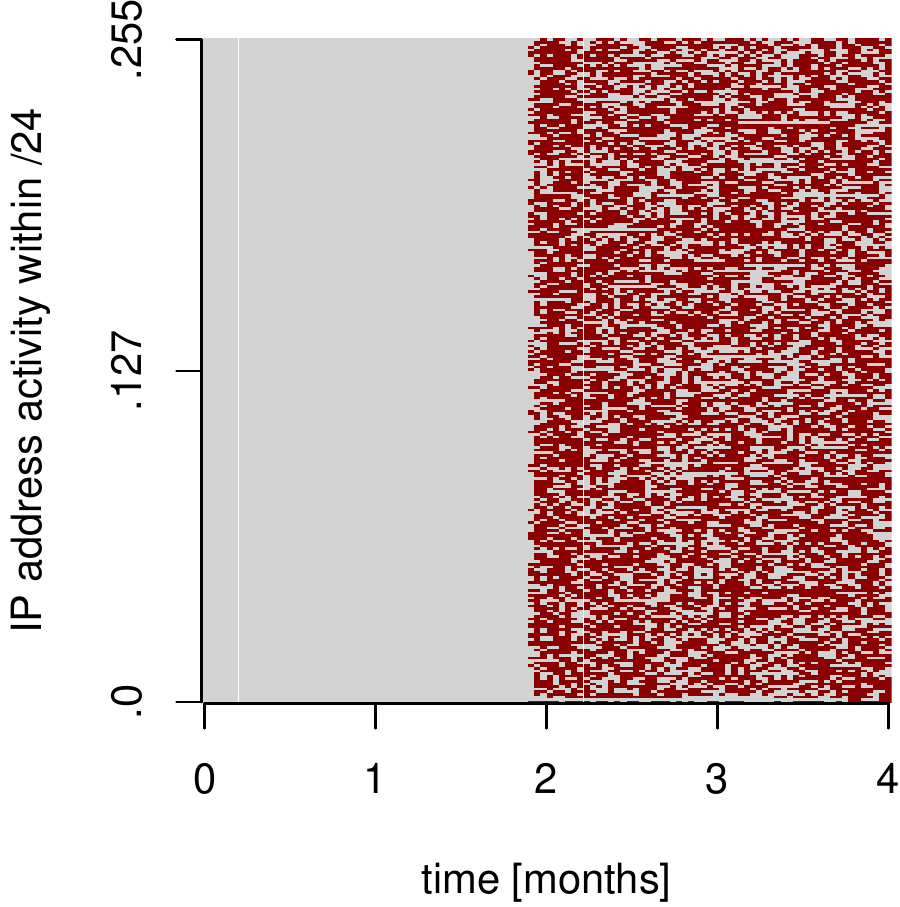}
    \label{fig:changes-population} %
  }
  \subfigure[German University, FD=187, STU=0.38.]{
    \includegraphics[width=0.47\linewidth]{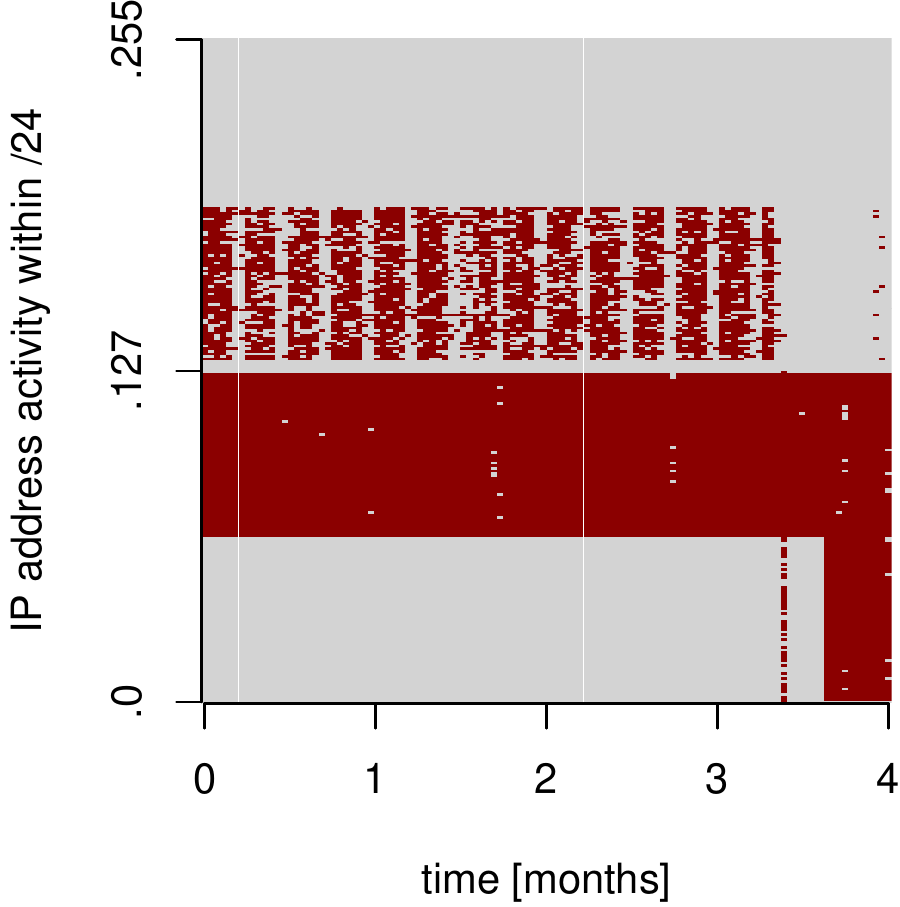}
    \label{fig:changes-administration} %
  }  
\caption{Modified assignment practice.}
\vspace{-1em}
\label{fig:microscopic-view-admin-change}
\end{figure}

\subsection{Block Activity Metrics}

In order to comprehensively characterize IP address activity, it is imperative to use metrics that capture the activity spatially, \ie over the IP address space of an address block, and temporally, \ie across time. %
To capture address activity patterns, next, we introduce two metrics:
\parax{IP address filling degree (FD)} this metric captures the number of 
active IPs within an address block within a window of time. There 
is not a single address block size that is ideal, but we chose a /24 Classless 
Inter-Domain Routing (CIDR) prefix, \ie the smallest distinct, globally-routed entity. 
This is a compromise, since we recognize that both smaller prefixes are 
sometimes more appropriate, as in Figure~\ref{fig:changes-administration}, and 
that larger prefixes sometimes exhibit uniform patterns of activity, e.g., 
Figure~\ref{fig:DHCP}. Values of this metric range from $1$ to $256$. We will 
later see that this metric is particularly helpful in dissecting static from 
dynamic addressing mechanisms.

\parax{Spatio-temporal utilization (STU)} this metric captures the aggregate activity of active IPs over time. We define utilization as the fraction: spatio-temporal activity divided by the maximum spatio-temporal activity, for a given block and window of observation (time). Relying on our four months worth (112 days) of daily activity data, the spatio-temporal activity can range from $1$, where one single IP address was active for one day, up to $112 \times 256 = 28672$, where all addresses in a block were active every day, which would be the maximum spatio-temporal activity. STU is this value normalized as a fraction with range 0 to 1. 

Figures~\ref{fig:microscopic-view} and~\ref{fig:microscopic-view-admin-change} are annotated with their respective values for filling degree (FD) and spatio-temporal utilization (STU). In these examples, FD varies from values as low as 29 to as high as 256. The STU varies from 0.04 up to 0.75.
\begin{figure*}[t]
  \subfigure[Maximum monthly change in spatio-temporal utilization per /24 block. We select 90\% of the blocks in the stable region.]{
    \includegraphics[width=0.315\linewidth]{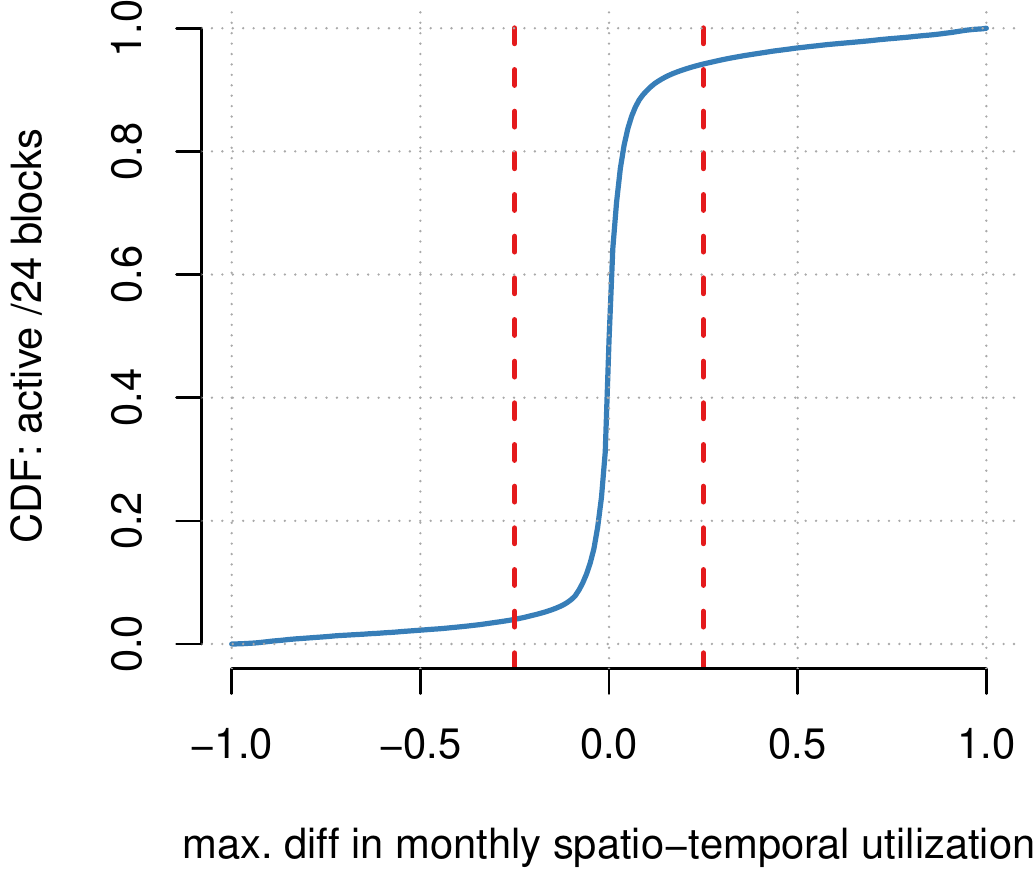}
    \label{fig:maxchange_spacetime}
  }
  \hfill
  \subfigure[Filling degree of active /24 blocks, where we dissect some identifiable blocks to be static or dynamic using reverse DNS.]{
    \includegraphics[width=0.315\linewidth]{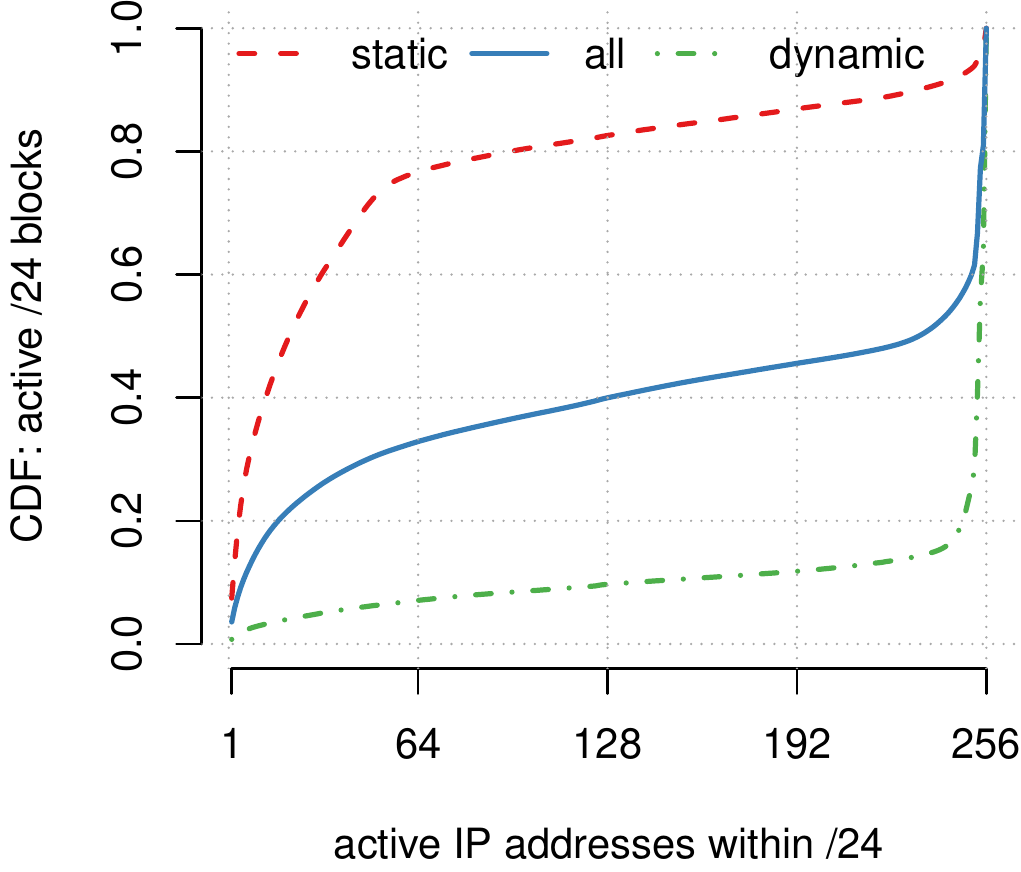}
    \label{fig:cdffilldegree}
  }
  \hfill
  \subfigure[Spatio-temporal utilization within /24 blocks with $>$250 active IP addresses.]{
    \includegraphics[width=0.315\linewidth]{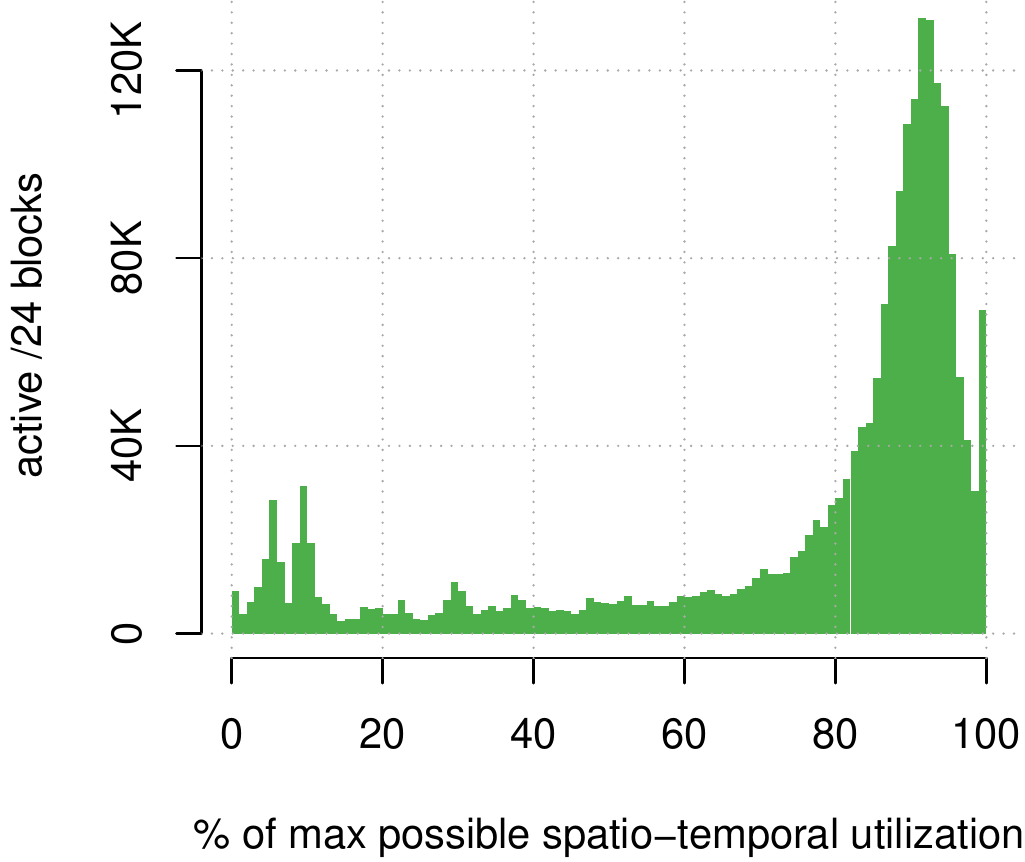}
    \label{fig:histutilization}
  }
  \hfill
\caption{Spatio-temporal aggregate views of IP address activity.}
\end{figure*}
\subsection{Detecting Change}
\label{sec:changeDetection}

As a first-order partitioning of the active IPv4 address space, we are interested in identifying address blocks with a significant change in address assignment practice during our observation interval. Per Section~\ref{sec:dissectingvolatility}, we know that some portion of address churn on longer timescales affects larger address ranges (``bulky changes'') than do short-term changes.  
To quantify changes in address assignment, we rely on our spatio-temporal 
utilization metric. In particular, Figure~\ref{fig:maxchange_spacetime} shows 
the maximum change in spatio-temporal utilization on a month-to-month  basis 
for each active /24 block. Here, we observe that the majority (90\%) of the /24 
blocks cluster around the origin, i.e., they do not show a major change in 
their utilization. Another 10\% of the active address blocks, on the other 
hand, are located more closely to the tails of the CDF, these are blocks for 
which we observe significant changes in address activity. 

To dissect address blocks into \textit{major change} and \textit{minor change} 
blocks, we set a threshold at $X=\pm0.25$. We decided to use this value, as it 
retains cases of heavy \textit{in situ} change, \eg Figure~\ref{fig:DHCP}, but 
excludes of major configuration change, \eg
Figure~\ref{fig:microscopic-view-admin-change}. Based on this threshold, we 
find that as many as 9.8\% of the active /24 blocks show major change in their 
address activity within our four months period, while 90.2\% of the blocks show 
no more than minor change. Thus, we separate blocks that likely underwent 
reallocation or change in address assignment practice (\textit{major change}, 
Figure~\ref{fig:microscopic-view-admin-change}) from those that did not 
(Figure~\ref{fig:microscopic-view}).\footnote{We acknowledge that some changes 
in address assignment might result in only minor STU change and that 
others might result in larger STU change. We 
chose a threshold based on anecdotal examination of activity patterns.}

\subsection{Addressing Matters}

Having culled out those blocks with major changes, we next focus on the activity characteristics of steady address
blocks. Since we have observed that the address assignment policy greatly influences its activity patterns, we would
like to identify specific assignment practices. We pay particular attention to \textit{utilization} characteristics
associated with these practices. We argue that an address block's utilization is determined by (a) its address assignment policy and (b) the behavior of its users and their hosts.

\parax{Static vs. dynamic addressing}
As a first cut, we are interested in how \textit{static} and \textit{dynamic} addressing mechanisms compare when it
comes to address space utilization. In the static case, the ISP assigns a fixed IP address for each device/subscriber.
Dynamic addressing, on the other hand, automatically assigns IP addresses from predefined ranges. In order to apply our
metrics, we wanted an initial set of blocks that are known to be likely statically or dynamically assigned. To this end, we used PTR
(reverse DNS) records and tagged /24 blocks containing addresses with consistent names that suggest static (keyword \texttt{static}) as
well as dynamic (keyword \texttt{dynamic, pool}) assignment, a well-known
methodology~\cite{How-Dynamic-IP-Addresses:SIGCOMM2007,quan2014internet,How-Dynamic-ISPs-Address:2015}.
In total, we find 456K dynamic /24 address blocks
and 262K static address blocks. We then compare their activity based on our 
dataset. Figure~\ref{fig:cdffilldegree} shows a CDF of the filling degree 
(active IPs per /24) for the two subsets of
static or dynamic /24s, as well as for the entirety of our dataset. Comparing the curves for dynamically and statically assigned
address blocks, we see a stark difference: While 75\% of static /24s show a filling degree lower than 64 IPs,
more than 80\% of the dynamic /24s show a very high filling degree, i.e., higher than 250 IP addresses.
When comparing these observations to our entire dataset, we observe that about 50\% of the entire
visible address space shows a very high filling degree (higher than 250). Another 30\%, by contrast, show
filling degrees lower than 64. If our DNS-derived samples are representative, most sparsely populated /24 blocks are statically assigned and
most dynamic pools cycle, i.e., have every address assigned at least once,
during our observation window of 4 months, resulting in a
high filling degree. However, about 20\% of the active /24s that remain have varying filling degrees. These
are either statically-assigned blocks with higher utilization or dynamically-assigned blocks with quite little
utilization, \eg those with long lease times as in
Figure~\ref{fig:churn_usisp}).

\begin{figure*}[t]
  \subfigure[Median daily hits per IP address binned by activity (days). The y-axis 
  is log-scaled.]{
    \includegraphics[width=0.30\linewidth]{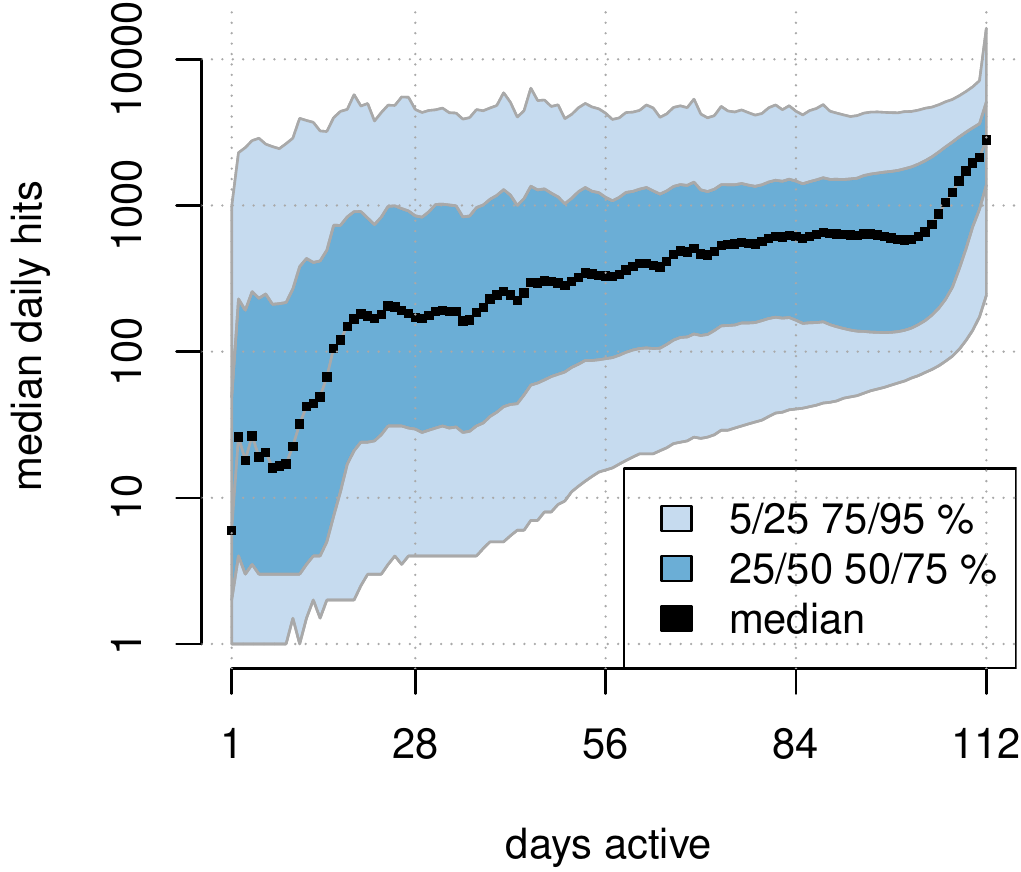}
    \label{fig:hits_daysactive_boxsplot}
  }
  \hfill
  \subfigure[Cumulative fraction of active IP addresses in each bin, cumulative 
  traffic contribution per bin.
  ]{
    \includegraphics[width=0.30\linewidth]{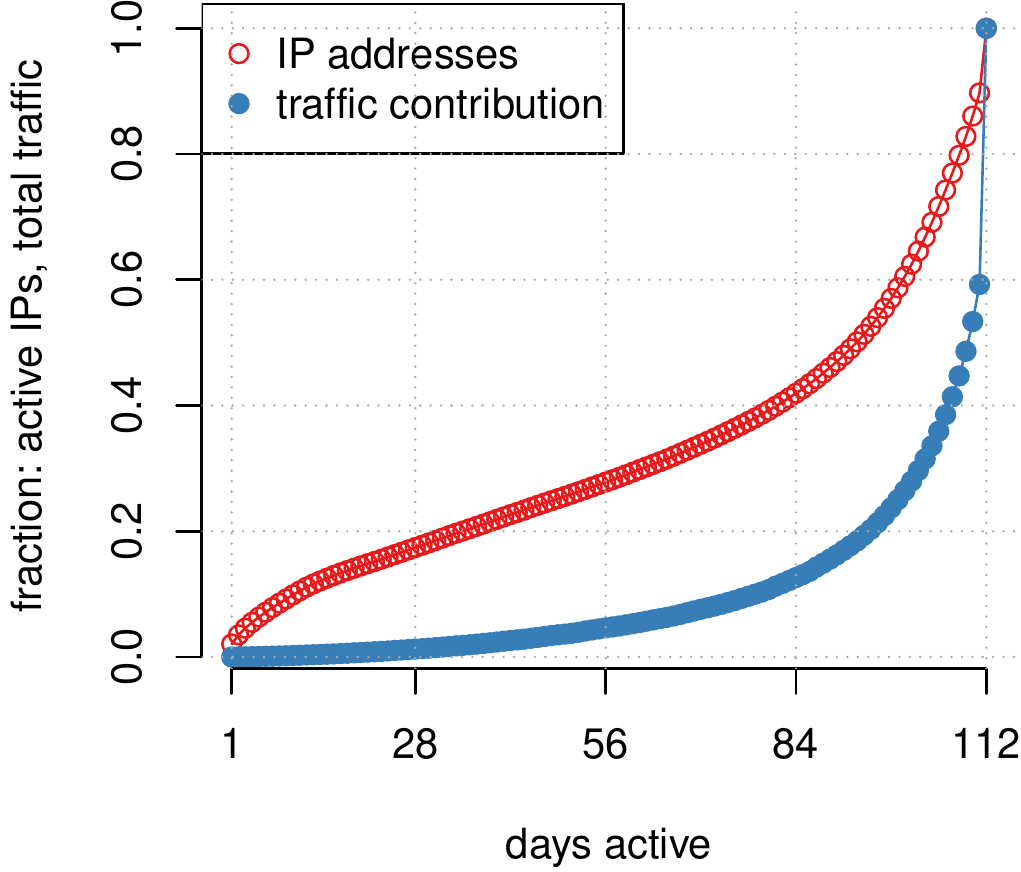} %
    \label{fig:daysactive_ips_traffic_cdf}
  }
  \hfill
  \subfigure[Relative share of total traffic of top 10\% IPs. The y-axis starts at 49\%.]{
    \includegraphics[width=0.30\linewidth]{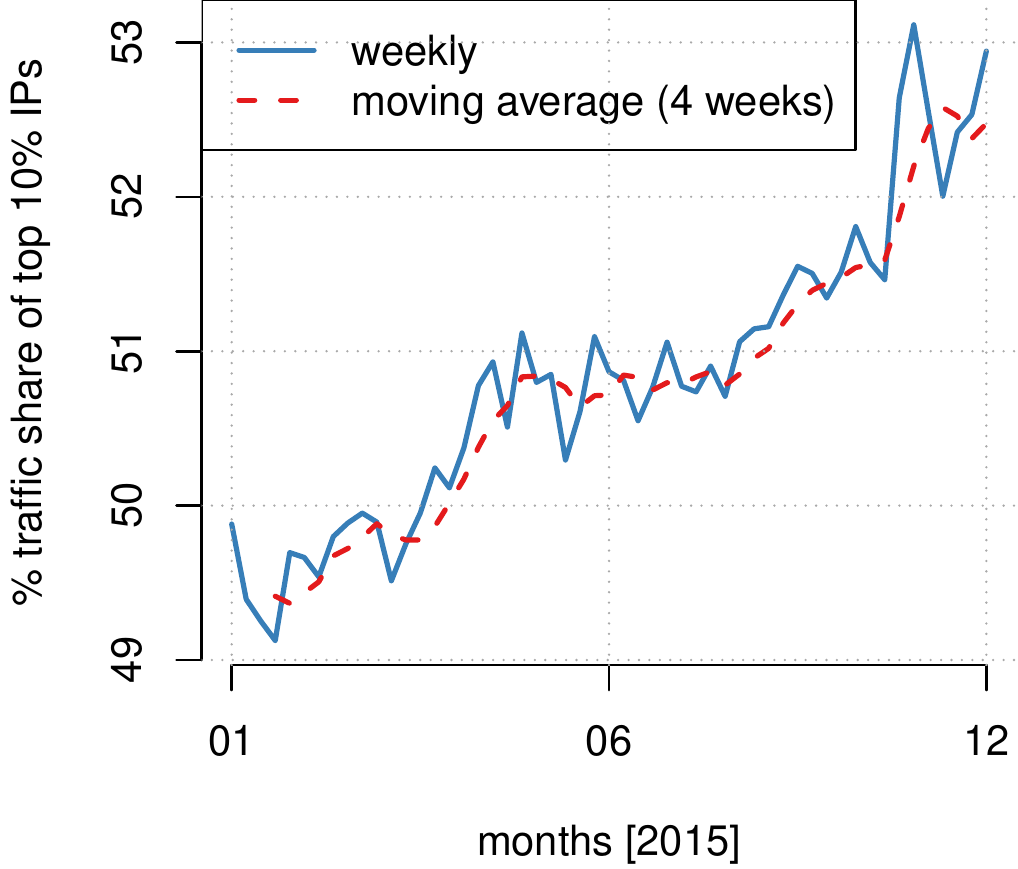}
    \label{fig:traffic_consolidation_trend}
  }
\label{fig:traffic_properties}
\caption{Activity time-range of IP addresses vs. their traffic contribution.}
\end{figure*}

\parax{Dynamic address pools}
We find that dynamically-assigned /24 prefixes generally show a very high 
filling degree, with more than 250 active IP addresses in more than 80\% of the 
cases. This heavily depends on the configured assignment policy, i.e., the 
\textit{pool size} in relation to the number of connecting devices. Figures 
\ref{fig:DHCP} and \ref{fig:dynamic-heavy} both show dynamic addressing 
patterns, however we see that their utilization is very different. To shed more 
light into such dynamic pools, we make use of our second metric, the 
\textit{spatio-temporal utilization}.  Focusing on those 1.2 million /24 blocks 
that have a very high filling degree (larger than 250, and hence likely 
dynamically assigned), Figure~\ref{fig:histutilization}, shows their 
spatio-temporal 
utilization as a percentage of their maximum possible 
utilization.\footnote{Figure~\ref{fig:histutilization} looks similar when only 
considering dynamic address blocks as identified using reverse DNS PTR 
records.}  Here, we see that most of these address blocks have high 
utilization, with most blocks at more than 80\%. In fact, we even see some 60K 
/24 blocks with 100\% spatio-temporal utilization. This extraordinary 
utilization hints that they might contain shared proxy or gateway addresses; we 
will revisit these in Section~\ref{sec:traffic-and-UAs}. We also see more than 
450K /24 prefixes with a utilization lower than 60\% and 200K /24s with a 
utilization even lower than 20\%.

\subsection{Potential Utilization}
Figure~\ref{fig:cdffilldegree} makes it clear that the spatio-temporal utilization of address blocks differs dramatically. We find that \textit{static} vs \textit{dynamic} addressing mechanisms play an important first-order role, and now present some estimates on an address block's maximum potential spatio-temporal utilization.  We constrain this exercise to only those blocks known to be active, i.e., those that are known to be allocated, globally-routed, and in operation.  We argue that increasing utilization in these blocks is -- {\em in some instances} -- a mere configuration issue.  Sometimes this means switching from static to dynamic assignment, but other times it means only reconfiguring an existing dynamic pool.

Specifically, we find that more than 30\% of the active IP address blocks,
more than 1.5M /24 blocks, have a filling degree lower than 64 active IP
addresses.
Our DNS PTR-based tagging method suggest that static address assignment practices are the main driver for low spatio-temporal utilization of IP address space.
On the other hand, for the 50\% of the active /24 address that appear to
be dynamically managed, we find that the majority have high spatio-temporal
utilization, i.e., more than 80\%. However, we also find that about one third 
of dynamic blocks show low spatio-temporal utilization; Figure~\ref{fig:DHCP} 
is a striking such example.  We argue that -- as these address blocks are 
already dynamically assigned -- reducing their pool sizes could instantly free 
significant portions of address space.  

\section{Traffic \& Devices}
\label{sec:traffic-and-UAs}

Up to this point in the paper, we studied the activity of an IPv4 address with respect to time and to neighboring addresses, \eg in a /24 prefix. We've seen a variety of address 
activity patterns and associated addressing mechanisms. Next, we take another dimension 
into account: traffic.  In particular, we'd like to answer these questions:
\textit{(i)} How 
does address activity correlate with traffic? \textit{(ii)} Do we see a 
long-term trend with respect to the fraction of traffic associated with the 
heavy-hitter addresses? \textit{(iii)} How does traffic contribution relate 
to the number of connected end hosts? Afterward, in 
Section~\ref{sec:clustering}, we will combine traffic metrics and host 
estimates with the activity measurements of Section~\ref{sec:microscopic} to 
obtain a comprehensive perspective of the active IPv4 address space.

\subsection{Activity vs. Traffic}\label{sec:traffic}

Firstly, we are interested in how the binary notion of acti\-vity of IP 
addresses is related to the volume of traffic that the CDN to delivers to them.
For this, we rely on our dataset that captures the number of daily HTTP 
requests as issued by each individual IP address (as described in 
Section~\ref{sec:cdn-perspective}).  We group all IP addresses 
that were active during our 4-month (112 days) period into 112 bins, 
corresponding to the number of days each individual IP address was active. 
Figure~\ref{fig:hits_daysactive_boxsplot} shows the median \textit{daily} hits 
that were issued by the total count of IP addresses in each bin, where we only 
consider days where an IP address issued at least one hit.  
We also show the 5, 25, 75 and 95 percentiles for each bin (the y-axis is 
log-scaled). Note the strong correlation between temporal activity of IP 
addresses and their daily traffic contribution. While addresses that were only active 
for a few days issue only a median of fewer than 100 requests per day, the 
traffic contribution is much higher for addresses that were active on more 
days. Indeed, we see that the traffic contribution significantly increases for 
IP addresses that were active almost every day ($\geq$ 110 days), and those 
addresses that were active every day show an even higher median daily traffic 
contribution.  This observation becomes clearer when looking at 
Figure~\ref{fig:daysactive_ips_traffic_cdf}, where we plot cumulative fractions of the total 
number of 
IP addresses falling into each bin (red) and their contribution to
the CDN client's total traffic (blue).
While only less than 10\% of IPv4 addresses were active every single day, these addresses account for more than 40\% of the CDN's total traffic!  The combination of continuous daily activity over the course of four consecutive months as well as the significantly larger contribution in overall traffic suggests that those 10\% of the active IPv4 addresses include gateways, e.g., NAT routers and web proxies, aggregating the traffic of multiple users, as well as WWW client bots (e.g., employed by search engines or content aggregators).

\subsection{Traffic Consolidation}

Given that we have reached a stage in which the number of active IPv4 addresses has stagnated, we were curious whether there is an
observable trend over 2015 of increasing traffic concentration in the 
heavy-hitter addresses. To visualize this, we show in 
Figure~\ref{fig:traffic_consolidation_trend} the traffic share of the 10\% of 
addresses with the greatest traffic.  (Note that the y-axis
starts at 49\%.)
Here, we use our weekly dataset to show how this trend has been developing over 
the entirety of the year 2015. Figure~\ref{fig:traffic_consolidation_trend} 
indeed shows a clear trend of traffic consolidation. 
While in January 2015, those IPv4 addresses received a share between 49\% and 50\%, we see that their traffic share steadily
increased over the course of the year.  As of December 2015, the top 10\% of the active IPv4 addresses consume an additional 3\% of the total 
traffic that the CDN serves, which we believe is a notable increase over one year. Given the stagnating count of active IPv4 addresses, we expect IPv4 traffic consolidation to continue except, \eg when and where alleviated by IPv6.

\subsection{Estimating Relative Host Counts}\label{sec:user-agents}

Having understood that the characteristics of activity of an IP address vary dramatically, both regarding its utilization as well as volume of traffic, we are next interested in how many hosts reside in a given address block.
With the increasing prevalence of address sharing mechanisms (e.g., Carrier-Grade NAT \cite{IMC2016-Richter-NATs}), active IP addresses are no longer an accurate metric to quantify the number of hosts in a given address block, e.g., to reason about Internet penetration, activity of individual users, or address activity in general.

While we do not have data available that provides us with a definitive number of connected hosts per IP address, we will estimate by \textit{HTTP User-Agent strings}, as a proxy. 
Whenever a Web object is requested from a server, the respective client application identifies itself by providing a User-Agent string within the HTTP request header. We extended the CDN data-collection platform to store a random sample of HTTP User-Agent strings of connecting hosts. Due to the high volume of this data, we only store the User-Agent field for 1 out of 4K HTTP requests, and we restrict this analysis to the last month of our observation period.

In the canonical case, the User-Agent identifies the browser version, OS version, as well as the screen resolution. However, in more recent times, primarily driven by smartphone applications, which typically identify themselves and their version number with an individual User-Agent string, we see much higher diversity in these strings~\cite{Xu:2011:IDU:2068816.2068847}.  %
HTTP User-Agent strings have been used in the past to quantify host populations behind NAT devices in residential networks \cite{maier2011nat}. Here, we use them only as a \textit{relative measure of host counts} per address block, i.e., we do not claim to be able to exactly quantify host populations. This is mainly because (a) the coarse-grained sampling of this dataset and (b) the fact that some single devices introduce multiple User-Agent strings (\eg a smartphone running many applications) while, simultaneously, those and other devices running the same applications might share an IP address which, thus, will consolidate \textit{unique} User-Agent strings (on a client address); the former can result in overestimation and the latter in underestimation of the host population.

\begin{figure}
    \centering
    \includegraphics[width=0.95\columnwidth]{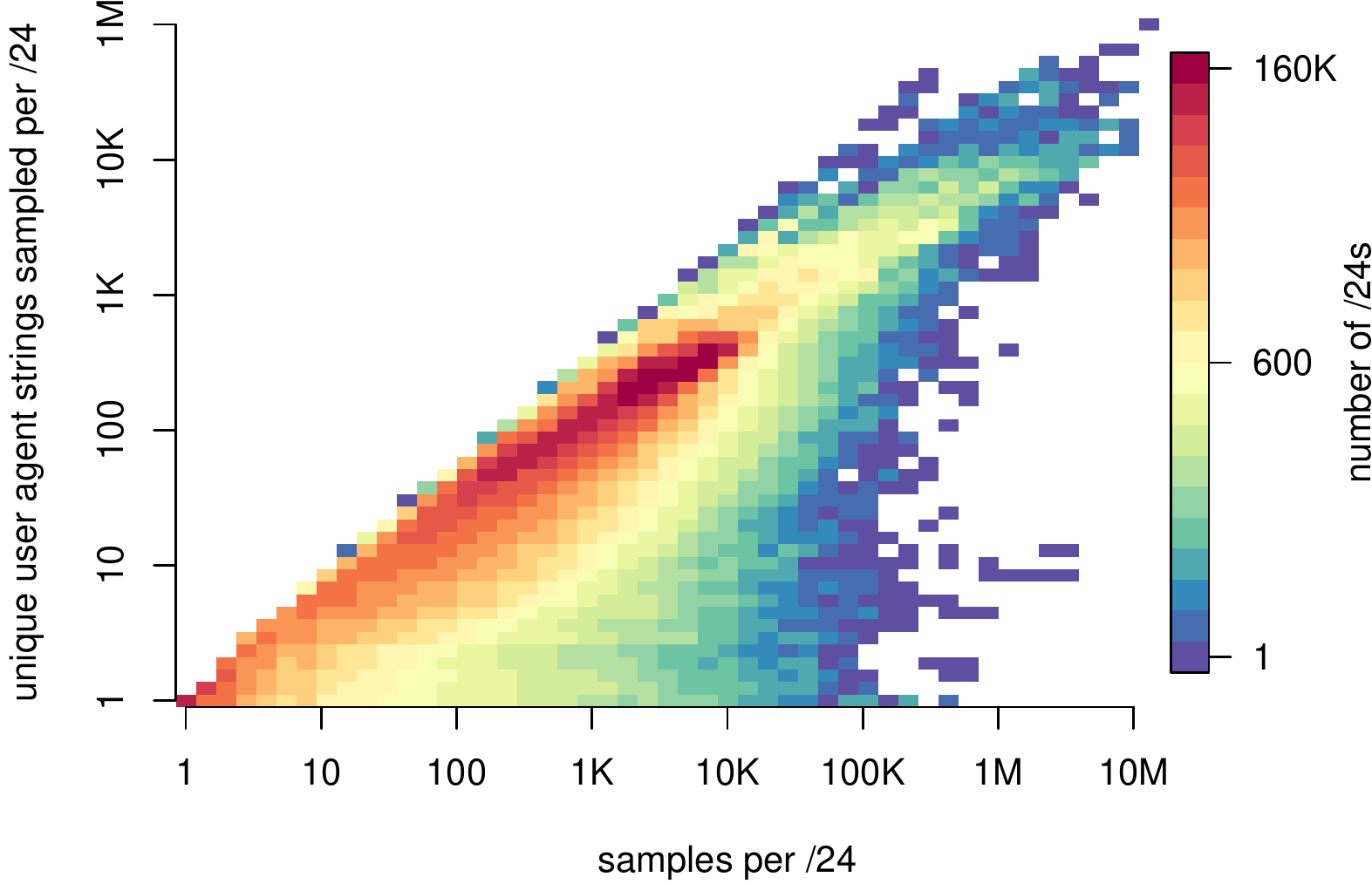}
    \caption{Diversity of User-Agent strings per /24 block.}
    \label{fig:scatter-hits-UAs}
\end{figure}

Figure~\ref{fig:scatter-hits-UAs} shows, for each active /24 block,
the number of User-Agent samples (x-axis) versus the number of \textit{unique} User-Agent 
Strings (y-axis). Thus, the x-axis value is an estimate of \textit{traffic volume} (based on sampled requests/hits) issued by hosts in a block and the y-axis value is a relative 
measure of the \textit{number of hosts} residing in a block. Overall, we see a 
strong correlation between traffic and hosts. Upon closer look, we can 
dissect the area in the plot in three groups: The first (and largest) group of 
/24 blocks ranges from the center of the figure to the lower left. Indeed, here 
we find the bulk of address blocks, e.g., from residential ISPs. Then, we have 
blocks that are shifted more towards the right, but show a low number of unique 
User-Agent strings (bottom right in the figure). By further investigation,
we found that these blocks are mainly related to automated activity, \eg
crawling bots, which 
issue a large number of requests, but do so with one (or very few) User-Agent 
string(s). More interestingly, we see a third region, in the top right, of a 
huge number of requests, and a very high diversity of User-Agent strings. A 
closer inspection of these blocks reveals that it is precisely those blocks 
that correspond to gateways, aggregating the traffic of thousands of end-users. 
We manually inspected the top 5K blocks in the top-right region of the plot. 
Using WHOIS information, we find that more than half of these blocks belong to 
ISPs located in Asia and that the majority is in use by cellular operators.

\section{Deriving Demographics}
\label{sec:clustering}

In this section we combine our activity metrics (spatio-temporal, traffic, relative host counts) to provide a comprehensive perspective of the active IPv4 address space. Our three features are fundamentally different in nature, which manifests itself also in different scaling of our derived values per address block. Hence, to project our features onto a unified scale, we first need to \textit{normalize} our measures of traffic and the relative host count. Our measure of spatio-temporal utilization is already normalized to a range $(0,1]$. We normalize the traffic contribution as well as the relative host count, by using a log-transform of the value per /24 block and divide it by the maximum log-transformed value of all active /24 blocks.
\begin{figure}
    \centering
    \includegraphics[width=.8\columnwidth]{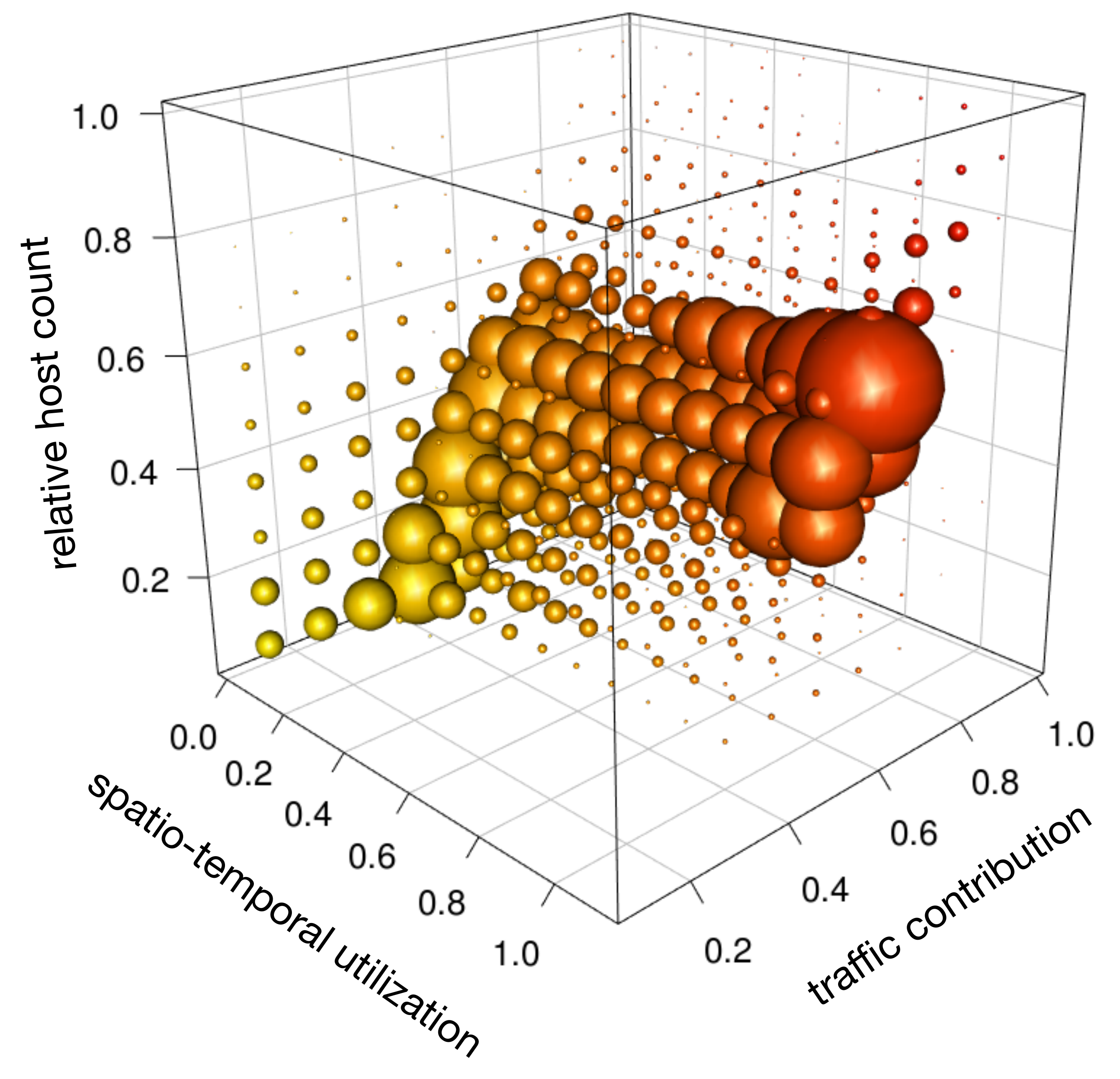}
    \caption{Characterization of the active IPv4 address space: Spatio-temporal activity, traffic contribution, relative host count per /24.}
    \label{fig:3d-clustering}
\end{figure}
Having these three normalized values per /24 block in hand, we next bin the resulting values into 10 intervals of a length of $0.1$. This results in a 3-dimensional array with $1000$ entries. We now assign each /24 block to one of these bins within our matrix.\footnote{For traffic contribution, the median of the 1st/5th/10th bin corresponds to $4$/$1.5M$/$44B$ monthly hits; For relative host density, the median of the 1st/5th/10th bin corresponds to $2$/$2K$/$500K$ unique sampled User-Agents strings.}

\subsection{Internet-wide Demographics}

Figure~\ref{fig:3d-clustering} shows a 3D-visualization of our feature matrix, 
where we indicate the number of /24 address blocks falling into each bin by 
scaling the size of the respective sphere. We can make several observations 
from this plot: \textit{(i)} We see a strong division of address blocks along 
the spatio-temporal utilization axis. While one set of blocks is clustered 
towards values with a very small spatio-temporal utilization (less than 0.2), 
another set is clustered towards very high spatio-temporal utilization. 
Recalling Section~\ref{sec:microscopic}, this can mainly be attributed to 
varying addressing mechanisms. \textit{(ii)} Taking the traffic 
contribution into account, we see that densely utilized address blocks 
typically have a higher traffic volume. However, this observation is not always 
true; we also see significant portions of the address space with high 
traffic volume in sparsely-populated areas. \textit{(iii)} When relating these 
two features to our host count measure, we again see a higher host count for 
highly-utilized and traffic-intensive blocks. In particular, we see only a very 
tiny portion of /24 blocks that fall into the highest bin for the host count 
metric. These blocks typically also show a maximum spatio-temporal utilization 
and maximum traffic contribution (small spheres at the top-right). It is 
important to notice that blocks contained in these small spheres are 
responsible for a significant share of the CDN's overall traffic.

\subsection{Regional Characteristics}

Lastly, we dissect the address space by regional registries. Recall that the address space is subject to management from 5 different organizations (RIRs, Section~\ref{sec:rethinking}). Each RIR applies different management policies and the current state of address exhaustion also varies per RIR. Thus, we believe that this grouping can assist in understanding the current status of the address space in each of these regions and support policy decisions when it comes to managing the last remaining blocks and re-allocations of address blocks already in use. Figure~\ref{fig:RIR} shows an address space categorization for the five RIRs. Here, we plot the spatio-temporal utilization and traffic contribution on the x and y axes, and indicate the relative host counts by the color scale (gray: low relative host count, red: high relative host count). Again, we adjust the size of the circles to reflect the number of /24s falling into each bin.

We can see that about half of the active address space within the ARIN region 
clusters towards the left, i.e., shows low utilization, low traffic 
contribution. However, we note that there are some heavily active address 
blocks also in this region (small red dots at x = 0.2 / y = 0.8,0.9). We see that the 
other regions have more of their address space being highly-utilized, 
which is especially true for LACNIC and AFRINIC. A possible explanation for 
this behavior is that LACNIC and AFRINIC were incorporated much later than the 
other RIRs and had address conservation as a primary goal from the very 
beginning~\cite{CCR-IPv4-scarcity}. Noticeably for 
the APNIC and AFRINIC regions, we see a significant chunk of /24 blocks 
towards the top-right of the figures (x = 1.0, y = 0.8), which also show a very 
high relative host count. This hints towards increased proxying/gateway 
deployments which is more pronounced in these regions when compared to, e.g., 
ARIN.

\begin{figure*}
    \centering
    \includegraphics[width=0.95\textwidth]{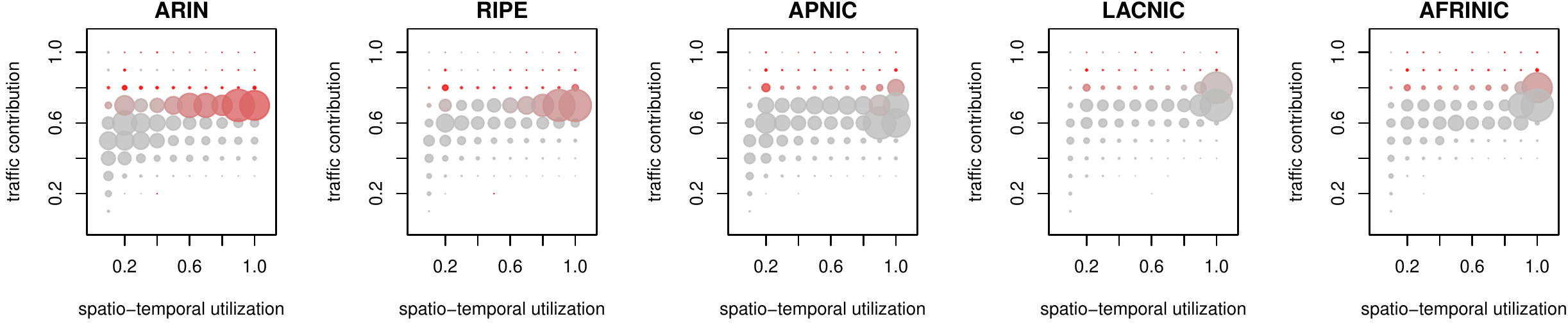}
    \caption{Breakdown of IP address space characterization per RIR. Color encodes the relative host count.}
    \label{fig:RIR}
\end{figure*}

\section{Implications}

\parax{Implications to measurement practice} 
We count 1.2 billion active, globally-unique IPv4 addresses, more than
has been reported previously, except by statistical
estimation~\cite{Capturing-Ghosts:IMC2014}, boding well for future use
of such statistical models and techniques driven by sampled observation.
Our address count analysis implies that remote
active measurements are insufficient for census or
complete survey of the Internet, particularly at IP address-level granularity.
Also, our passive measurements have shown extensive churn in IPv4 addresses on all timescales,
which implies that any census needs to be qualified by the observation frequency and period.

\parax{Implications to Internet Governance}
The 1.2 billion active addresses we count represent 42.8\% of the possible
unicast addresses that we see advertised in the global routing table. If we restrict our
implications to the 6.5 million /24 prefixes in which we observed active
WWW clients addresses (Table~\ref{tab:dataset}), \ie exclude blocks that may be
dedicated to network infrastructure and services, we see that roughly
450 million address may have been unused. 
If some large subset actually
are unused, today, one could imagine reallocating them for use in IPv6
transition mechanisms that require IPv4 addresses, \eg NAT64 and
DNS64~\cite{RFC6146,RFC6147}, or as a commodity whose supply might last
years in a marketplace, based on past rates of growth in IPv4 address
use (Figure~\ref{fig:active-IPv4-evolution}).  

IPv4 address markets are an operational reality, governed by the respective RIR policies~\cite{CCR-IPv4-scarcity}. A pertinent implication of our work for these markets is that our metrics, combined with the appropriate vantage points, are ideal to readily determine spatio-temporal utilization of network blocks. This can aid RIRs in determining the current state of address utilization in their respective regions, in determining if a transfer conforms with their transfer policy (e.g., four of five RIRs require market transfer recipients to justify need for address space), as well as in identifying likely candidate buyers and sellers of addresses. 

\parax{Implications to network management}
It is feasible for any network to employ our metrics and perform
our analysis on a continual basis, \eg by monitoring traffic at its border.
Measuring spatio-temporal utilization would enable an operator to 
more efficiently manage the IPv4 addresses they assign, especially in
networks such as those discussed in Section~\ref{sec:traffic-and-UAs}.
Networks that make gains in efficiency by discovering unnecessary address blocks
may decide to become sellers in the IPv4 transfer marketplace. More generally, we believe that our measurements can serve as input for fruitful discussions on address assignment practices and their eventual effect on address space utilization. 

\parax{Implications to network security}
Our observations of many, disparate rates of change in the assignment of IP
addresses to users has consequences for maintaining host-based access controls
and host reputations.  A host's IP address is often associated with a
reputation subsequently used for network abuse mitigation,
\eg in the form of access control lists and application rate-limits that
specifically use those IP network blocks or addresses as identifiers with
which some level of trust is (or is not) associated.
Unfortunately, in this way, addresses and the network blocks 
become encumbered by their prior uses and the behavior of users within.
This happens when reputation
information is stale. %
The implication of our work here is that it can {\em inform} host-based
access control and host reputation, \eg by determining the spatial and
temporal bounds beyond which an IP addresses reputation should no longer
be respected.  Further, our change detection method
(Section~\ref{sec:changeDetection}) could be used to trigger expiration
of host reputation, avoiding security vulnerabilities
when networks are renumbered or repurposed.

\parax{Implications to content delivery}
Details about user activity at the address level are valuable
in CDN operation. A key responsibility of CDNs is to map users to the
appropriate server(s) based on criteria including performance and
cost~\cite{Akamai-Network}.
Details about active IP addresses and network blocks are 
increasingly important when the CDN uses end-user
mapping~\cite{Akamai-Mapping-EDNS-2015},
where client addresses are mapped to the appropriate server.

\section{Conclusion}

In this paper, we study the Internet through the lens of IPv4 address-level
activity as measured by successful connections to a large CDN.
After many years of constant growth, active IPv4 address counts have stagnated,
while IPv6 counts have grown~\cite{plonka2015temporal}.
In addition, we observe churn in the set of active addresses on time scales 
ranging from a day to a year. %
Simple address counts do not capture the increasingly complex situation of usage of the IPv4 address space.
Instead, we use three metrics that our results
show are helpful to understand what is happening now:
spatio-temporal aspects of address activity,
address-associated traffic volume, and relative host counts.
Continued overall growth but lagging adoption of IPv6 have brought a
reimagined IPv4 upon us, one that entails increased address sharing in both space
and time.
The Internet community is in a complex and costly resource-limited
predicament, foreseen but unavoided.
The prolonged tussle continues amongst operators about whether and when to
implement incremental changes to IPv4, adopt IPv6, or both.  
Our study, as well as others that might adopt our metrics, can guide us in this 
tussle and better illuminate the condition of the IPv4 address space.

\section*{Acknowledgments}

This work would not have been possible without the full support of the Akamai 
Custom Analytics team. In particular we thank Matt Olson, Keung-Chi ``KC'' Ng, 
and Steve Hoey. We thank the anonymous reviewers for their useful comments and 
suggestions. Georgios Smaragdakis was supported by the EU Marie Curie IOF 
``CDN-H'' (PEOPLE-628441). This work was partially supported by Leibniz Prize 
project funds of DFG - German Research Foundation (FKZ FE 570/4-1).

\bibliographystyle{plain}
\bibliography{paper}
\end{document}